\documentclass{article}
\pdfoutput=1

\usepackage{natbib}
\usepackage{epubtk}
\usepackage{amsmath}
\usepackage{amssymb}
\usepackage{graphicx}
\usepackage{textcomp}
\usepackage{xspace}

\showlistoftablesfalse

\newcommand{\kms}{km~s\super{-1}\xspace}
\newcommand{\ms}{m~s\super{-1}\xspace}
\newcommand{\Mxcm}{Mx~cm\super{-2}\xspace}

\begin{document}

\title{The Sun's Global Photospheric and Coronal Magnetic Fields: Observations and Models}

\author{
\epubtkAuthorData{Duncan H.\ Mackay}{
University of St Andrews \\
School of Mathematics and Statistics, University of St Andrews, \\
North Haugh, St Andrews, Fife, KY16 9SS, UK}{
duncan@mcs.st-and.ac.uk}{%
http://www-solar.mcs.st-andrews.ac.uk/~duncan}%
\and
\epubtkAuthorData{Anthony R.\ Yeates}{%
Department of Mathematical Sciences \\
Durham University, Science Laboratories, \\
South Road, Durham DH1 3LE, UK}{%
anthony.yeates@durham.ac.uk}{%
http://www.maths.dur.ac.uk/~bmjg46}%
}

\date{}
\maketitle

\begin{abstract}

In this review, our present day understanding of the Sun's global photospheric and coronal
magnetic fields is discussed from both observational and theoretical viewpoints. 
Firstly, the large-scale properties of photospheric magnetic fields are described, along with 
recent advances in photospheric magnetic flux transport models.
Following this, the wide variety 
of theoretical models used to simulate global coronal magnetic fields are described. From this,
the combined application of both magnetic flux transport simulations and coronal modeling techniques 
to describe the phenomena of coronal holes, the Sun's open magnetic flux and the hemispheric pattern of 
solar filaments is discussed. Finally, recent advances in non-eruptive global MHD models are described. 
While the review focuses mainly on solar magnetic fields, recent advances in measuring and modeling 
stellar magnetic fields are described where appropriate. In the final section key areas of future research 
are identified. 
\end{abstract}

\epubtkKeywords{Magnetic fields, Photosphere, Corona, observations, modelling}

\newpage
\tableofcontents

\newpage

\section{Introduction}
\label{sec:intro}

Magnetic fields play a key role in the existence and variability of a wide variety of 
phenomena on the Sun. These range from relatively stable, slowly evolving 
features such as sunspots \citep{2003AARv..11..153S}, coronal loops \citep{2010LRSP....7....5R}, and solar prominences
\citep{2010SSRv..151..243L,2010SSRv..151..333M}, to highly dynamic phenomena 
such as solar flares \citep{2008LRSP....5....1B} and coronal mass ejections 
\citep{2006SSRv..123..251F,2011LRSP....8....1C}. Solar magnetic fields may directly or indirectly 
affect the Earth through the Sun's open magnetic flux \citep{1995Sci...268.1007B}, solar wind
\citep{2008JApA...29..217H}, and total solar irradiance variations \citep{1998GeoRL..25.4377F}. 
 
Our present day understanding of solar magnetic fields dates back to 1908 when G.E.~Hale made 
the first magnetic field observations of sunspots \citep{1908ApJ....28..315H}. However, it was 
not until the systematic mapping of the Sun's magnetic field carried out by the Babcocks
\citep{1952PASP...64..282B,1958Sci...127.1058B,1959ApJ...130..364B} that the true nature of solar
magnetic activity became apparent. 
While significant advances in observations have been made over the last 50 years, only the strength 
and distribution of the line-of-sight component at the level of the photosphere has been regularly 
measured over solar cycle time-scales. However, over the last 5 years, vector magnetic field measurements
at the photospheric level have been systematically carried out by the satellite 
missions of Hinode since 2006 \citep{2007SoPh..243....3K} and SDO (Solar Dynamics Observatory) since 2010 \citep{2012SoPh..275....3P}. In addition, systematic ground based
measurements of vector magnetic fields have been made by SOLIS (Synoptic Optical Long-term Investigations of the Sun) since 2009 \citep{2003SPIE.4853..194K}. While Hinode only observes 
vector magnetic fields over localised areas, the global capabilities of SDO and SOLIS provide us with a unique 
opportunity to observe the large-scale spatial distribution of vector magnetic fields across the solar surface. 
Over time scales of years, SDO and SOLIS will significantly enhance our understanding of the emergence and transport of magnetic
fields at the level of the photosphere and their subsequent effect on the global corona.
While magnetic fields may be observed directly at the photospheric level, due to low densities
the same is not true for the corona. As many important 
phenomena occur in the solar corona, a key component in our understanding of solar 
magnetic fields -- and the build up of energy within them -- is the use of theoretical models to construct 
(or extrapolate) coronal magnetic fields from photospheric data.  

While solar magnetic fields have been observed in detail over long periods of time, the same is not true for other
stars. In recent years however, the technique of Zeeman Doppler Imaging \citep{1989AA...225..456S} has
lead to a significant advance in our understanding of magnetic fields on other stars. 
Results show a wide range of magnetic distributions across stars of varying mass and spectral 
type \citep{2009AIPC.1094..130D}. With the accurate measurement of stellar 
magnetic fields, techniques developed to model solar magnetic fields, both at photospheric and coronal levels, are now widely applied 
in the stellar context \citep{2010ApJ...721...80C,2011IAUS..273..242J}.

In this review, we primarily focus on our present day understanding of global solar magnetic 
fields from both observations and theoretical models. The reader should note that we focus on quasi-static or steady-state coronal models and will not consider fully dynamic eruptive models. 
Within the review we also restrict 
ourselves to global aspects of the Sun's magnetic field so will consider neither
the Sun's small-scale field nor limited field-of-view models. The review will focus in particular
on advances made in the last 15 years, although we have not aimed at completeness in material or 
references. Additional topics will be added in future revisions. 
Where appropriate, we expand this discussion into stellar magnetic fields 
to summarise new results or to describe the application of models developed for the Sun in the 
stellar context. The review is split into three distinct parts, where each part is largely self-contained. 
Thus the reader may focus on each part separately if desired. The review is split in the following 
way:

\newpage
\begin{itemize}

\item \textbf{Global Photospheric Magnetic Fields (Section~\ref{sec:sec1})} \\
Observations of global solar and stellar 
photospheric magnetic fields are briefly discussed (Section~\ref{sec:obs}). Following from this, 
magnetic flux transport models used to simulate the spatial and temporal evolution of photospheric 
magnetic fields are described, along with the variety of extensions and applications of these models
(Section~\ref{sec:mftm}). For a historical discussion of magnetic flux transport models, the reader is referred to the article by \citet{2005LRSP....2....5S}. 

\item \textbf{Global Coronal Models (Section~\ref{sec:cf})} \\ 
This section of the review surveys the wide variety of 
techniques used to model global coronal magnetic fields. This includes the various approximations
that are applied from static extrapolation techniques, to time dependent quasi-static models
and finally recent advances in global non-eruptive MHD models.

\item \textbf{Application of Global Models (Section~\ref{sec:app})} \\
The final part of the review considers the combined application of both 
magnetic flux transport models and coronal modeling techniques to model a variety of
phenomena found on the Sun. These include the Sun's open magnetic flux (Section~\ref{sec:open}),
coronal holes (Section~\ref{sec:ch}), and the hemispheric pattern of solar filaments (Section~\ref{sec:fil}).
In addition, recent advances in MHD models for modeling the plasma 
emission of the corona are also considered (Section~\ref{sec:mhd}). 
\end{itemize}
Finally, in Section~\ref{sec:con} a brief summary is given and some outstanding problems or areas of advancement are
outlined.

\newpage

\section{Photospheric Magnetic Fields}
\label{sec:sec1}

\subsection{Observations}
\label{sec:obs}

Presently, three solar cycles of continuous data have been collected
by a variety of ground- and space-based observatories (Mount Wilson
Observatory, Wilcox Solar Observatory, Kitt Peak, SoHO/MDI, SOLIS)
mapping the distribution and evolution of the Sun's normal magnetic
field component at the level of the photosphere. An illustration of
this can be seen in Figure~\ref{fig:mackay1}a
\citep[from][]{2010LRSP....7....1H}. The image is known as the solar
magnetic ``butterfly diagram'' and illustrates the longitude-averaged
radial magnetic field as a function of time (horizontal axis) versus
sine-latitude (vertical axis). Yellow represents positive flux and
blue negative flux, where the field saturates at \textpm~10~G. The
main features in the long-term evolution of the global magnetic field
are:
\begin{itemize}

\item[-] At the start of each solar cycle (ca.\ 1975, 1986, 1996) the
  majority of magnetic flux sits in the polar regions which are of
  opposite polarity. New magnetic flux then emerges in the form of
  sunspots or large magnetic bipoles in two latitude bands between
  \textpm~30\textdegree\ \citep{1913MNRAS..74..112M}. As the cycle
  progresses these bands approach the equator.

\item[-] As new flux emerges in the form of magnetic bipoles, the polarities lie mainly in an east-west orientation. In each hemisphere the leading polarity (in the direction of rotation) has the same sign as the polar field of the hemisphere in which it lies, the following polarity has the opposite sign \citep[Hale's Polarity Law,][]{1925ApJ....62..270H}. In addition, the majority of bipoles emerge subject to Joy's Law, where the leading polarity lies equator-ward of the following \citep{1919ApJ....49..153H, 1991SoPh..136..251H}. The effect of Joy's Law is clearly seen in Figure~\ref{fig:mackay1}a as a latitudinal separation of magnetic fluxes: following polarity dominates at latitudes poleward of \textpm~30\textdegree; leading polarity dominates at latitudes closer to the equator.

\item[-] This latitudinal separation combined with the effects of
  meridional flow \citep{1979SoPh...63....3D,2010Sci...327.1350H} and
  the dispersal of magnetic flux out from complexes of activity, leads
  to the preferential transport poleward of the following polarity. In
  contrast, the leading polarity in each hemisphere, which lies at
  lower latitudes, partially escapes the effect of meridional flow to
  disperse and cancel across the equator. As a result, in each
  hemisphere more following flux is transported poleward, where it
  first cancels the existing polar field and then builds up a new
  polar field of following polarity. This transport of flux poleward
  occurs in quasi-episodic poleward streams which extend out of the
  active latitudes. Such poleward streams are anchored in closely
  packed clusters of sunspots. The clusters may persist in the same
  location for several solar rotations as new bipoles emerge to
  refresh the cluster. These sites are referred to as complexes of
  activity \citep{1983ApJ...265.1056G} or as activity nests
  \citep{2000ssma.book.....S}. \cite{1993SoPh..148...85H} found that
  at least half of the active regions larger than 3.5~deg\super{2}
  emerge within nests. Nests themselves are sometimes grouped together
  as interacting units; the grouping format determines how flux
  streams poleward as the groups decay \citep{2008ApJ...686.1432G}.

\item[-] The reversal in sign of the polar field typically occurs
  1\,--\,2~years after cycle maximum. In recently observed cycles this
  process occurs with every 11-year activity cycle, with a further
  11-year activity cycle required before the polar fields in each
  hemisphere return to their initial sign. A 22-year magnetic cycle
  therefore overlies the 11-year activity cycle.

\end{itemize}

\epubtkImage{mackay_fig2.png}{%
\begin{figure}[htb]
\centering\includegraphics[scale=1.0]{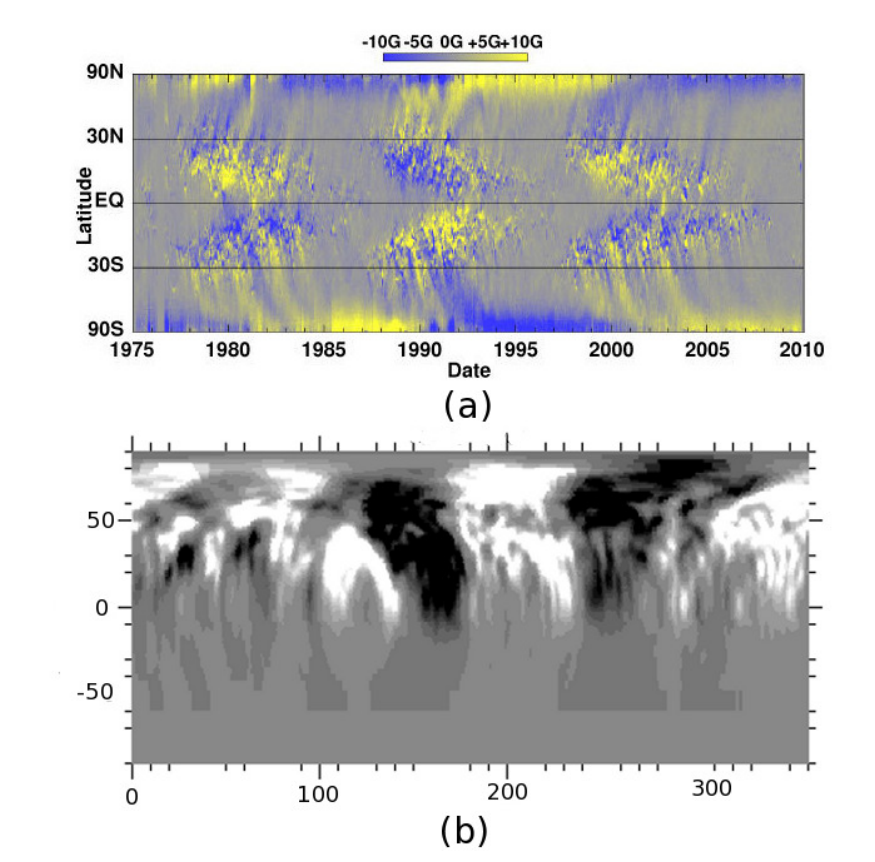}
\caption{(a)~The Solar Butterfly Diagram
  \citep[reproduced from][]{2010LRSP....7....1H}. Yellow represents
  positive flux and blue negative flux where the field saturates at
  \textpm~10~G. (b)~Example of a typical radial magnetic field
  distribution for AB~Dor taken through Zeeman Doppler Imaging
  \citep[ZDI, data from][]{2003MNRAS.345.1145D} where the image
  saturates at \textpm~300~G. White/black denotes positive/negative
  flux. Due to the tilt angle of AB~Dor, measurements can only be made
  in its northern hemisphere. Image reproduced by permission
  from~\cite{2004MNRAS.354..737M}, copyright by RAS.}
\label{fig:mackay1}
\end{figure}}

While continuous global measurements of magnetic activity only exist from around
1975, observations of the numbers of sunspots may be used to provide a long running
data set of solar activity back to 1611 \citep{1998SoPh..181..491H,2011ApJ...731L..24V}. These
show that on top of the approximate 11 or 22-year activity cycle there are strong modulations 
in the number of sunspots (or magnetic flux emergence rate) over periods of centuries. 
It is also possible for large-scale magnetic activity to disappear. Such an event
occurred between 1645 and 1715 where it is known as the Maunder 
minimum \citep{1976Sci...192.1189E,1993AA...276..549R}.
Before 1611 indicators of magnetic activity on the Sun may be found through
the use of proxies such as \super{14}C \citep{1980Sci...207...11S} and \super{10}Be isotopes \citep{1990Natur.347..164B}. 
Through this, reconstructions of the level of magnetic activity over the past 10,000 years may be made 
\citep{2008LRSP....5....3U} and show that many such ``grand minima'' have occurred over the last 10,000
years.

While our present day knowledge of solar magnetic fields is vast, the majority of this knowledge 
comes from observing the line-of-sight component at the level of the photosphere. 
To gain a much fuller understanding of the Sun's magnetic field,
vector field measurements are required \citep{1994SoPh..155....1L}. However, these measurements 
are complicated to make, with problems including low signal-to-noise ratios and resolving
the 180\textdegree\ ambiguity. In addition, in the past such measurements have not been regularly 
made until the launch of Hinode (ca.\ 2006) which can make vector magnetic field measurements over
localised areas of the Sun. However, with the new space mission of SDO (launched in 2010) and the availability of
ground based vector magnetic field measurements (SOLIS), such measurements 
should now be  made regularly over the full solar disk in strong field regions. This -- combined with modeling techniques -- should 
significantly enhance our understanding of solar magnetic fields in years to come. 
While current  vector magnetic field measurements should increase our knowledge, to fully understand the Sun's magnetic 
field vector, measurements are also required in weak field regions over the entire Sun. This poses a significant technical 
challenge, but is something that future instrument designers must consider.

While our understanding of magnetic fields on stars other than the Sun is at an early stage,
significant progress has been made over the last 10 years. For young, rapidly rotating solar-like stars, 
very different magnetic field distributions may be found compared to the Sun. An example of this can be 
seen in Figure~\ref{fig:mackay1}b, where a typical radial magnetic field distribution for AB~Dor taken 
through Zeeman Doppler Imaging \citep[ZDI,][]{1989AA...225..456S} is shown. AB~Dor has a rotation period 
of around 1/2~day, which is
significantly shorter than that of the Sun (27~days).
Compared to the Sun, key differences include kilogauss polar fields covering a large area of the pole 
and the mixing of both positive and negative polarities at the poles. While this is an illustration 
of a single star at a single time, many such observations 
have been made across a wide range of spectral classes. In Figure~3 of \citet{2009AIPC.1094..130D} 
the varying form of morphology and strength of the magnetic fields for a number of stars  ranging in 
spectral class from early F to late M can be seen compared to that of the Sun. The plot covers
stars with a rotation period ranging from 0.4 to 30~days and masses from (0.09 to $2\,M_{\odot}$). While
ZDI magnetic field data sets are generally too short to show cyclic variations, recent observations
of the planet-hosting star, $\tau$ Bootis, have shown that it may have a magnetic cycle with period of only
2 years \citep{2009MNRAS.398.1383F}. Indirect evidence for cyclic magnetic field variations on other stars 
can also be seen from the Mt.~Wilson Ca\,{\scriptsize II}~H+K observations, which use chromospheric observations as a proxy for 
photospheric magnetic activity \citep{1995ApJ...438..269B}. These show that magnetic activity on stars
of spectral types G2 to K5V has three main forms of variation. These are (i) moderate activity and
regular oscillations similar to the Sun, (ii) high activity and irregular variations (mainly seen
on young stars), and finally (iii) stars with flat levels of activity. The final set are assumed
to be in a Maunder like state. In the next section magnetic 
flux transport models used to simulate the evolution of the radial magnetic field at the 
level of the photosphere on the Sun and other stars are discussed.

\subsection{Magnetic flux transport simulations}
\label{sec:mftm}

On large spatial scales, once new magnetic flux has emerged on the Sun, it evolves through the advection processes 
of differential rotation \citep{1983ApJ...270..288S} and meridional flow 
\citep{1979SoPh...63....3D,2010Sci...327.1350H}.  In addition, small convective cells such as super-granulation 
lead to a random walk of magnetic elements across the solar surface. On spatial scales much larger than super-granules this random walk may 
be modeled as a diffusive process \citep{1964ApJ...140.1547L}. Magnetic flux transport simulations 
\citep{2005LRSP....2....5S} apply these effects to model the large-scale, long-time evolution of the radial 
magnetic field $B_r(\theta,\phi,t)$ across the solar surface. In Section~\ref{sec:basicmftm} the basic formulation 
of these models is described. In Section~\ref{sec:extmftm} 
extensions to the standard model are discussed and, finally, in Sections~\ref{sec:appmftm1}\,--\,\ref{sec:appmftm4}
applications of magnetic flux transport models are considered.

\subsubsection{Standard model}
\label{sec:basicmftm}

The standard equation of magnetic flux transport arises from the
radial component of the magnetic induction equation under the
assumptions that $v_r= 0$ and $\partial/\partial r=0$.%
\epubtkFootnote{Alternatively, the magnetic flux transport equation may
  be obtained through spatially averaging the radial component of the
  induction equation (see \citealp{1984SoPh...92....1D} and
  \citealp{2002SoPh..211...53M}).}
These assumptions constrain the radial field component to evolve on a
spherical shell of fixed radius, where the time evolution of the
radial field component is decoupled from the horizontal field
components. Under these assumptions, the evolution of the radial
magnetic field, $B_r$, at the solar surface ($R_{\odot}=1$) is
governed by
\begin{equation}
\frac{\partial {B}_{r}}{\partial {t}} = \frac{1}{\sin \theta} 
\frac{\partial}{\partial \theta}\left(\sin\theta\left( -{u}(\theta){B}_r 
+{D}\frac{\partial {B}_r}{\partial \theta}\right)\right)  - 
\Omega(\theta)\frac{\partial {B}_r}{\partial \phi} +  
\frac{{D}}{ \sin^2 \theta} \frac{\partial^2 {B}_r}{\partial \phi^2} + S(\theta,\phi,t),
\label{eq:ft}
\end{equation}
where $\Omega(\theta)$ and $u(\theta)$ represent the surface flows of differential rotation and  meridional flow, 
respectively, which passively advect the field,
$D$ is the isotropic diffusion coefficient representing superganular diffusion, and finally $S(\theta,\phi,t)$ is an 
additional source term added to represent the emergence of new magnetic flux. Figure~\ref{fig:testft} illustrates two numerical solutions to the flux transport equation when $S=0$. Both are initialised with a single bipole in the northern hemisphere, which is then evolved forward in time for 30 solar rotations. The simulations differ only in the tilt of the initial bipole: in the left column, the bipole satisfies Joy's law, while in the right column both polarities lie at the same latitude. The effect
of Joy's law has a significant impact on the strength and distribution of $B_r$ across the surface of the Sun and, in particular,
in the polar regions.

A wide range of studies have been carried out to determine the best
fit profiles for the advection and diffusion processes
\citep[see][]{1985SoPh..102...41D, 1985AuJPh..38..999D,
  1989Sci...245..712W, 1998ApJ...501..866V}. The parameter study of
\citet{2004AA...426.1075B} demonstrates the effect of varying many of
the model parameters. The most commonly accepted values are the
following:

\epubtkImage{mackay_surface.png}{%
\begin{figure}[htbp]
\centering\includegraphics[width=\textwidth]{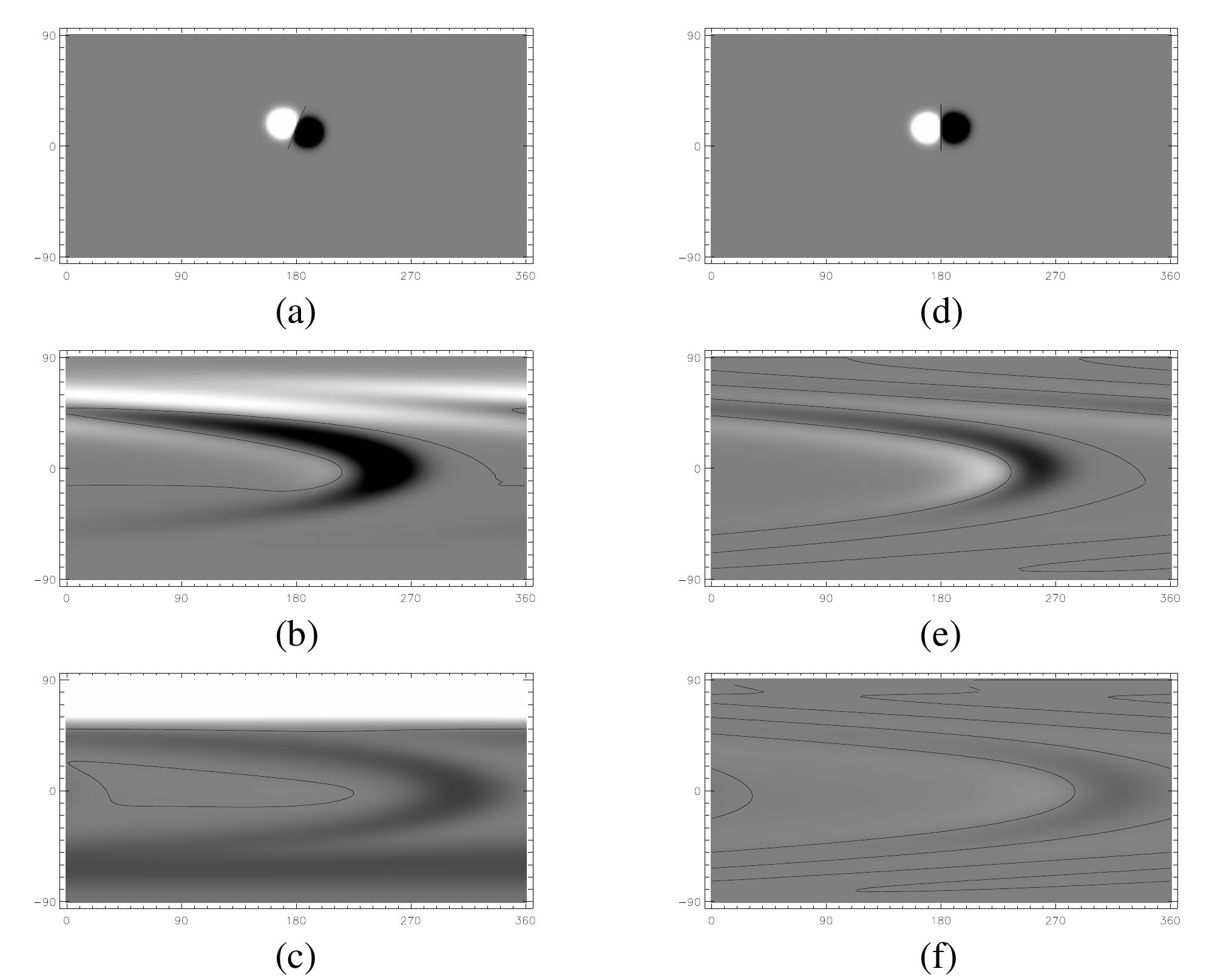}
\caption{Evolution of the radial component of the magnetic field ($B_r$) at 
the solar surface for a single bipole in the northern hemisphere with
(a)\,--\,(c) initial tilt angle of $\gamma$~=~20\textdegree\ and (d)\,--\,(f)
$\gamma$~=~0\textdegree. The surface distributions are shown for (a)
and (d) the initial distribution, (b) and (e) after 15 rotations, and
(c) and (f) after 30 rotations. White represents positive flux and
black negative flux and the thin solid line is the Polarity Inversion
Line. The saturation levels for the field are set to 100~G, 10~G, and
5~G after 0, 15, and 30 rotations, respectively.}
\label{fig:testft}
\end{figure}}

\begin{enumerate}
\item \textbf{Differential Rotation:} 
The form that best agrees with the evolution of magnetic flux seen on
the Sun \citep{1985SoPh..102...41D} is that of
\cite{1983ApJ...270..288S}. The profile (Figure~\ref{fig:ftpro}a) is
given by
\begin{equation}
\Omega(\theta) = 13.38 - 2.30 \cos^2 \theta - 1.62 \cos^4 \theta \mathrm{\ deg/day},
\label{eqn:omega}
\end{equation}
and was determined by cross-correlation of magnetic features seen on
daily Mt.~Wilson magnetogram observations. The key effect of
differential rotation is to shear magnetic fields in an east-west
direction, where the strongest shear occurs at mid-latitudes (see
Figure~\ref{fig:ftpro}a). This produces bands of alternating
positive and negative polarity as you move poleward (visible after 15
rotations in Figure~\ref{fig:testft}). These bands result in steep
meridional gradients which accelerate the decay of the
non-axisymmetric%
\epubtkFootnote{The evolution of the axisymmetric (longitude-averaged)
  $B_r$ is independent of the differential rotation, as may be seen by
  averaging Equation~\eqref{eqn:omega} in longitude \citep{1964ApJ...140.1547L}.} 
field during periods of low magnetic activity
\citep{1987SoPh..112...17D}. On the Sun the timescale for differential
rotation to act is $\tau_{\mathrm{dr}}= 2\pi/(\Omega(0)-\Omega(90)) \sim 1/4$
year. Within magnetic flux transport simulations the
profile~\eqref{eqn:omega} is either applied directly, simulating the
actual rotation of the Sun, or with the Carrington rotation rate
(13.2~deg/day) subtracted
\citep{1998ApJ...501..866V,2002SoPh..211...53M}.

\epubtkImage{mackay_fig7.png}{%
\begin{figure}[htbp]
\centering\includegraphics[width=\textwidth]{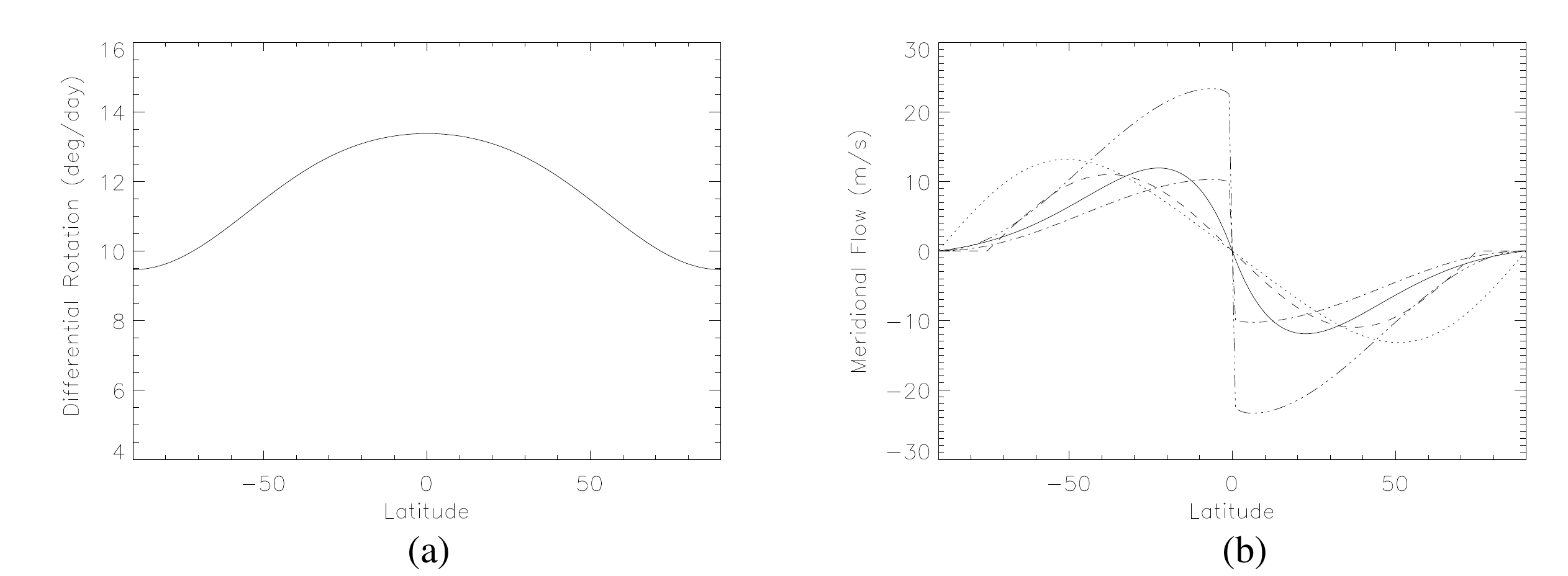}
\caption{(a)~Profile of differential rotation $\Omega$ versus latitude
  \citep{1983ApJ...270..288S} that is most commonly used in magnetic
  flux transport simulations. (b)~Profiles of meridional flow $u$
  versus latitude used in various studies. The profiles are from
  \cite{2006AA...459..945S} (solid line), \cite{1998ApJ...501..866V}
  (dashed line), \cite{2001ApJ...547..475S} (dotted line), and
  \cite{2002ApJ...580.1188W} (dash-dot and dash-dot-dot-dot
  lines). This figure is based on Figure~3 of
  \cite{2006AA...459..945S}.}
\label{fig:ftpro}
\end{figure}}

\item \textbf{Supergranular Diffusion:} 
This represents the effect on the Sun's large-scale magnetic field of the  
small-scale convective motions of supergranules. The term was first introduced by \cite{1964ApJ...140.1547L} 
to represent the effect of the non-stationary pattern of supergranular cells dispersing magnetic flux 
across the solar surface. In addition it describes the cancellation of magnetic flux (see Figure~\ref{fig:testft}) 
when positive and negative magnetic elements encounter one-another. Initial estimates based on obtaining the 
correct reversal time of the polar fields 
(without including meridional flow) were of a diffusion coefficient of
$D \sim 770\mbox{\,--\,}1540\mathrm{\ km^{2}\ s^{-1}}$. 
However, once meridional flow was included to aid the transport of magnetic flux poleward, the value was lowered 
to around $200\mbox{\,--\,}600\mathrm{\ km^{2}\ s^{-1}}$ which better agrees with observational estimates and is commonly used in models today 
\citep{1985SoPh..102...41D,1989Sci...245..712W}. Globally this gives a time-scale  
$\tau_{\mathrm{mf}}= R_{\odot}^2/D \sim 34\mbox{\,--\,}80\mathrm{\ yr}$, however when considering non-global length scales such as that of
individual active regions the timescale is much shorter. Some models have introduced a discrete random walk process as an alternative to the diffusion term which will be discussed in Section~\ref{sec:extmftm} 
\citep[e.g.,][]{1994ApJ...430..399W,2001ApJ...547..475S}.

\item \textbf{Meridional Flow:} 
This effect was the last to be added to what is now known as the standard magnetic flux transport model. It 
represents an observed weak flow that pushes magnetic flux from the equator to the
poles in each hemisphere. The exact rate and profile applied varies
from author to author, but peak values of 10\,--\,20~\ms are commonly used. Figure~\ref{fig:ftpro}b shows a number of meridional flow profiles that have been used by different authors. The typical time-scale for 
meridional flow is $\tau_{\mathrm{mf}}= R_{\odot}/u(\theta) \sim 1\mbox{\,--\,}2\mathrm{\ yr}$. As current measurements of meridional flow are 
at the limits of detection, for both the flow rate and profile, this has lead to some variations in the exact
profile and values used which will be discussed in more detail in Section~\ref{sec:appmftm2}.
The first systematic study of the 
effects of meridional flow on the photospheric field was carried out by \cite{1984SoPh...92....1D}. This showed 
that to obtain a 
realistic distribution and strength for the polar field, the meridional flow profile must peak at low to 
mid-latitudes and rapidly decrease to zero at high latitudes. The inclusion of meridional flow was a key 
development which meant that much lower rates of the diffusion coefficient could be applied, while still 
allowing the polar fields to reverse at the correct time. It also aids in producing the observed ``topknot'' latitudinal profile of the polar field \citep[more concentrated than a dipole,][]{1989SoPh..124....1S}, and in reproducing the strong poleward surges observed in the butterfly diagram \citep{1989ApJ...347..529W}. In recent years more significant changes to the profile
of meridional flow have been suggested. This will be discussed in Section~\ref{sec:appmftm2} and 
Section~\ref{sec:appmftm3}.

\item \textbf{Magnetic Flux Emergence:} The final term in Equation~(\ref{eq:ft})  is a time-dependent
source term which represents a contribution to the radial magnetic field from the emergence of new 
magnetic bipoles. Most flux transport simulations carry out emergence in a semi-empirical way where the
emergence is carried out instantaneously, so that
growth of the new bipole is not considered.  Instead, only its decay under the action of
the advection and diffusion process described above are followed. Inclusion of this term is critical
in simulations extending over more than one rotation, to ensure that accuracy of the photospheric field is maintained. 
Within the literature the source term has been specified in a number of ways and  reproduces
the main properties of the butterfly diagram, along with varying levels of magnetic activity through
single cycles and from one cycle 
to the next. Different ways in which the source term has been
specified include:
\begin{enumerate}
\item \textbf{Observationally determining the properties of new bipoles from daily or synoptic magnetograms} 
so that actual magnetic field configurations found on the Sun may be reproduced 
\citep{1985SoPh...98..219S,2007SoPh..245...87Y}.
Statistical variations of these properties have been applied to model multiple solar cycles of varying activity.

\item \textbf{Producing synthetic data sets from power law distributions} \citep{1993SoPh..148...85H,1994SoPh..150....1S}   
where the power laws specify the number of bipoles emerging at a 
given time with a specific area or flux  \citep{1998ApJ...501..866V,2001ApJ...547..475S}. 

\item \textbf{Using observations of sunspot group numbers to specify the number of bipoles emerging} 
\citep{2002ApJ...577.1006S,2006AA...446..307B,2010ApJ...709..301J} where the flux within the bipoles 
can be specified through empirical sunspot area-flux relationships \citep{2006AA...446..307B,2010ApJ...709..301J}.
Recently \cite{2011AA...528A..82J} extended this technique back to 1700 using both 
Group \citep{1998SoPh..181..491H} and Wolf \citep{1861MNRAS..21...77W} sunspot numbers.

\item \textbf{Assimilating observed magnetograms directly into the flux transport simulations} 
\citep{2000SoPh..195..247W,2003ApJ...590..493S,2004SoPh..222..345D}. \citet{2000SoPh..195..247W} use the flux transport model to produce evolving magnetic synoptic maps from NSO Kitt Peak data. They develop a technique where full-disk magnetograms 
are assimilated when available and a flux transport model is used to fill in for unobserved or poorly observed regions 
(such as the far side of the Sun or the poles).
\cite{2003ApJ...590..493S} use a similar technique but with full-disk MDI observations, demonstrating it over a 5~yr period.
For the near 
side of the Sun, the MDI observations are inserted every 6~h to reproduce the actual field over a 60\textdegree\ 
degree window
where the measurements are most accurate. The authors show that the magnetic flux transport 
process correctly predicts the return of magnetic elements from the far side, except for the case 
of flux that emerged on the far-side. To account for this, the authors also include far side 
acoustic observations for the emergence of new regions \citep{2000SoPh..192..261L,2001ApJ...560L.189B}. 
In contrast to the method of \cite{2003ApJ...590..493S}, which inputs 
observations from daily MDI magnetograms, \cite{2004SoPh..222..345D} inserted fields from synoptic magnetograms
once per solar rotation for all latitudes between \textpm~60\textdegree. They used this to investigate the transport of flux poleward
and the reversal of the polar field. Recently, the \citet{2000SoPh..195..247W} model has been incorporated in a more rigorous data assimilation framework to form the Air Force Data Assimilative Photospheric Flux Transport Model \citep[ADAPT,][]{2010AIPC.1216..343A,2012SpWea..1002011H}. An ensemble of model realisations with different parameter values allow both data and model uncertainties to be incorporated in predictions of  photospheric evolution.

\end{enumerate} 

A common treatment of the source term is to include only large magnetic bipoles of
flux exceeding 10\super{20}~Mx. Extensions of the model to include small-scale magnetic regions are described in Section~\ref{sec:extmftm}.
While many of the parameters for newly emerging bipoles
may be determined observationally, or specified through empirical relationships, the parameter about which there is most disagreement in the literature is the variation of the tilt angle ($\gamma$) with latitude, or Joy's law. While it should in principle be observed directly for each individual magnetic bipole, this is possible only in recent cycles for which magnetogram observations are available. Traditionally the tilt angle was chosen to vary with latitude $\lambda$ as 
$\gamma \sim \lambda/2$, but more recent studies suggest that a much smaller variation with latitude is required \citep[$\gamma \sim 0.15 \lambda$,][]{2006AA...459..945S}. The tilt angle is a critical quantity as variations can have a significant effect on the net amount of magnetic flux pushed poleward (see Figure~\ref{fig:testft}) and subsequently on the reversal times of the polar fields and amount of open flux. This will be discussed further in Sections~\ref{sec:appmftm2} and \ref{sec:open}. 
\end{enumerate}

\subsubsection{Extensions}
\label{sec:extmftm}

Since the early flux transport models were produced \citep{1985SoPh..102...41D}, new variations have 
been developed with new formulations and features added by a variety of authors. These include:
\begin{enumerate}
\item Extending the basic flux transport model for the large-scale field to include the emergence
of small-scale fields down to the size of ephemeral regions \citep{2000SoPh..195..247W,2001ApJ...547..475S}. 
Such small scale fields are a necessary element to model the magnetic network and simulate chromospheric
radiative losses. Including the small scale emergences leads to a much more realistic description of the
photospheric field where discrete magnetic elements can be seen at all latitudes (compare Figure~\ref{fig:schr}
to Figure~\ref{fig:testft}), however since they are randomly oriented small-scale regions do not have a significant 
effect on large-scale diffusion or on the polar field \citep{1991ApJ...375..761W,2000SoPh..195..247W}. To 
resolve such small-scales, \citet{2001ApJ...547..475S} introduced a particle-tracking concept along with rules for the interaction
of magnetic elements with one-another. Another important extension
introduced by \cite{2001ApJ...547..475S} was  magneto-convective coupling. With its introduction the early decay 
of active regions could be considered, where strong field regions diffuse more slowly due to the
suppression of convection.

\item Introducing an additional decay term into the basic flux transport model \citep{2002ApJ...577.1006S,2006AA...446..307B}.
In the paper of \citet{2006AA...446..307B} this term is specified as a linear 
decay of the mode amplitudes of the radial magnetic field. The term 
approximates how a radial structure to the magnetic field, along with volume diffusion, would affect the radial
component that is constrained to lie on a 
spherical shell. The authors only include a decay term based on the lowest order radial mode, 
as it is the only radial mode with a sufficiently long decay time to affect the global field. Through 
simulating the surface and polar fields from Cycles~13\,--\,23 (1874\,--\,2005) and requiring the simulated polar 
fields to reverse at the correct time as given by both magnetic and proxy observations, the authors
deduce a volume diffusivity of the order of $50\mbox{\,--\,}100\mathrm{\ km^{2}\ s^{-1}}$. 

\item 
Coupling the magnetic flux transport model to a coronal evolution model so that both the 
photospheric and coronal magnetic fields evolve together in time. The first study to consider this was 
carried out by \cite{1998ApJ...501..866V}. Limitations of this initial model were that in the coronal volume the 
radial ($B_r$) and horizontal ($B_\theta,B_\phi$) field components evolve independently from one 
another, and that no force balance was considered for the coronal
field. Later developments \citep[][see Section~\ref{sec:ftmf}]{1998ApJ...501..866V,2000ApJ...539..983V,2006ApJ...641..577M}
removed all of these restrictive assumptions so that a fully coupled evolution of the coronal field 
along with radial diffusion, radial velocities, and force balance occurred. In an additional study the 
technique was also extended to include a simplified treatment of the convective zone \citep{2007ApJ...659.1713V}.

\item Reformulating the flux transport equation into a ``synoptic''
  transport equation, that evolves synoptic magnetic field maps such
  as those produced by NSO/Kitt Peak%
\epubtkFootnote{\url{http://nsokp.nso.edu/}} 
from one Carrington rotation to the next \citep{2002SoPh..211...53M}.
Their synoptic transport equation takes the form
\begin{equation}
\frac{\partial {\mathcal B}_{r}}{\partial {\tau}} = 
-\frac{f(\theta)}{\sin \theta} \frac{\partial}{\partial \theta}\left(\sin\theta {u}(\theta){\mathcal B}_r\right)
- \frac{v_\phi(\theta) f(\theta)}{\sin\theta}\frac{\partial {\mathcal B}_{r}}{\partial \phi}
+ \frac{Df(\theta)}{\sin \theta}\frac{\partial}{\partial \theta}\left( \sin \theta \frac{\partial {\mathcal B}_r}{\partial \theta}\right)
+ \frac{Df(\theta)}{\sin^2 \theta}\frac{\partial^2 {\mathcal B}_r }{\partial \phi^2}, 
\label{eq:sft}
\end{equation}
where ${\mathcal B}_{r}$ represents the synoptic magnetogram, $\tau$ a time index such that integer values of
$2\pi/\Omega$ correspond to successive synoptic magnetograms, and $f(\theta) = 1/( 1+ v_\phi(\theta)/(\Omega \sin \theta))$. The individual terms in Equation~(\ref{eq:sft}) are similar to those of Equation~(\ref{eq:ft}) but
modified to take into account that, due to differential rotation, different latitudes return to 
central meridian at different times.

\item Although the surface flux transport model is primarily considered in 2D, \citet{1964ApJ...140.1547L} also introduced 
a reduced 1-D form where the large-scale radial field is spatially averaged in the azimuthal direction. Due to the form of 
Equation~\eqref{eqn:omega}, the differential rotation $\Omega(\theta)$ plays no role in such a model. Recent applications 
of this 1-D model have been to reconcile surface flux transport with results from axisymmetric, kinematic dynamo simulations 
in the $r$-$\theta$ plane. \citet{2007ApJ...659..801C}, through describing the
emergence of magnetic flux based on sunspot records in a similar manner to that used by \cite{2006GeoRL..3305102D} and 
\cite{2006ApJ...649..498D}, determine whether the 1-D surface flux transport model has any predictive properties. They show that 
in some cases a significant positive correlation can be found between the amount of flux canceling across the equator
and the strength of the next cycle. However, when detailed observations of bipole emergences are added, they find that
any predictive power may be lost. \citet{2012AA...542A.127C} have used the same model to derive constraints on parameter values in the dynamo models, based on observed flux evolution at the surface.

\end{enumerate}

\epubtkImage{mackay_fig8.png}{%
\begin{figure}[htbp]
\centering\includegraphics[scale=0.4]{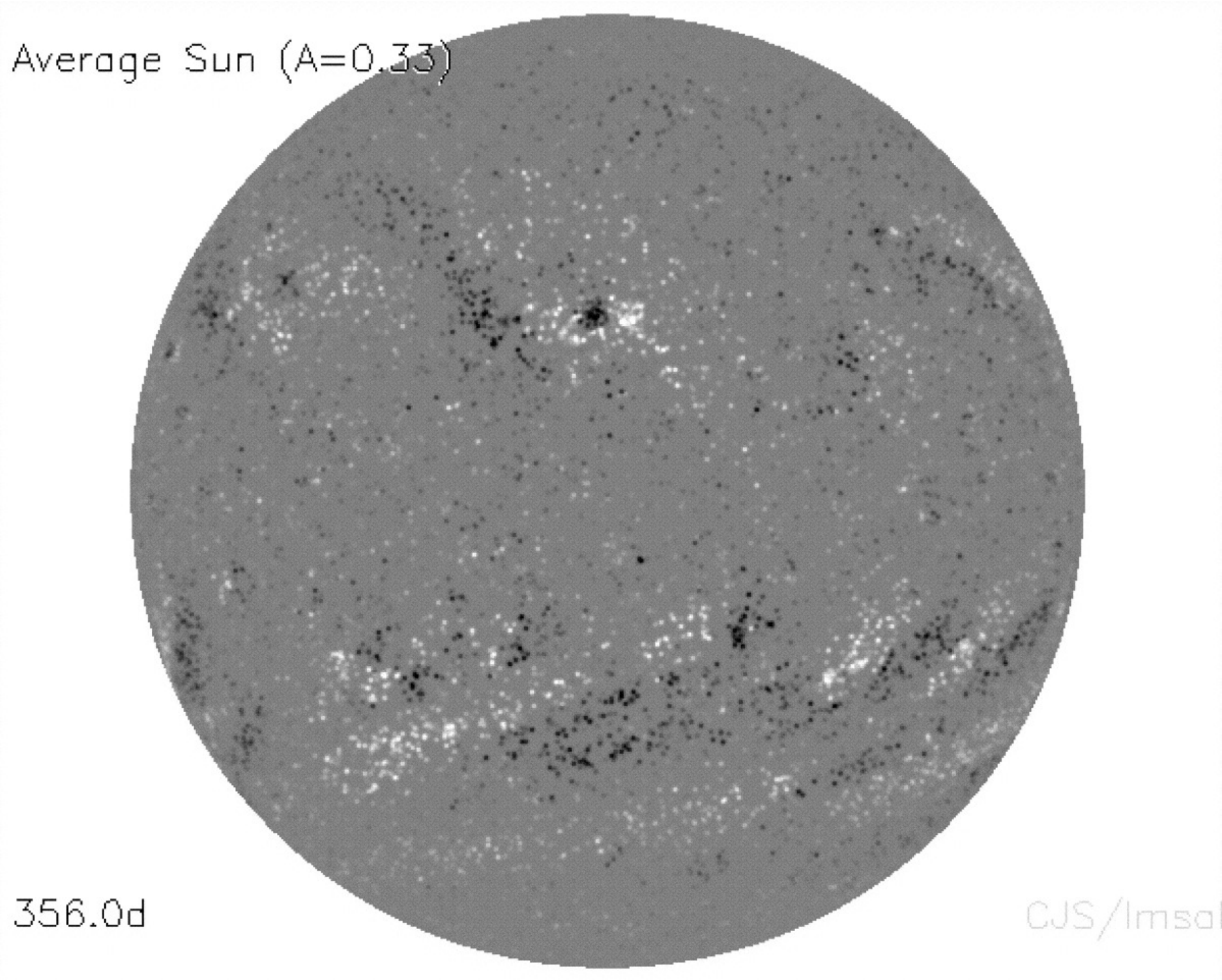}
\caption{Example of the magnetic flux transport simulations of
  \citet{2001ApJ...547..475S} which apply a particle-tracking concept
  to simulate global magnetic fields down to the scale of ephemeral
  regions. Image reproduced by permission
  from~\cite{2001ApJ...547..475S}, copyright by AAS.}
\label{fig:schr}
\end{figure}}

The magnetic flux transport models described above have been widely used and extensively compared with 
observations of the Sun's large-scale magnetic field. These comparisons are described next for time-scales of
up to a single 11-year cycle (Section~\ref{sec:appmftm1}) and also for multiple solar cycles 
(Section~\ref{sec:appmftm2}). In addition, simulations where the profile of meridional flow is changed from that 
shown in Figure~\ref{fig:ftpro}b are considered
(Section~\ref{sec:appmftm3}) along with the application of these models to stellar magnetic fields 
(Section~\ref{sec:appmftm4}).

\subsubsection{Short term applications of magnetic flux transport models}
\label{sec:appmftm1}

It is generally found that for simulations extending up to a full 11-year solar cycle (termed here short term
simulations) magnetic flux transport models are highly successful in reproducing 
the key features of the evolution of the Sun's radial magnetic field.  These include:
\begin{enumerate}
\item The 27~day and 28\,--\,29~day recurrent patterns seen in plots of the Sun's mean line-of-sight field.
These patterns can be respectively attributed to differential rotation and the combined effect of 
differential rotation with magnetic flux emergence \citep{1985SoPh...98..219S}.

\item  Reproducing the rigid rotation of large-scale magnetic features at the equatorial rate, caused by a balance 
between differential rotation and a poleward drift due to the combined effects of meridional 
flow and surface diffusion \citep{1987SoPh..112...17D,1987ApJ...319..481S,1989Sci...245..712W}.

\item Reproducing the reversal time of the polar fields along with the strength and top-knot
profile \citep{1987SoPh..108...47D,2003SoPh..214...23D}. 

\item Matching qualitatively the strength and distribution of the radial magnetic field at the
photosphere. Examples include the formation of switchbacks of polarity inversions lines 
at mid-latitudes due to the dispersal of magnetic flux from low latitude active regions 
\citep{1989Sci...245..712W,1998ApJ...501..866V} and the return of magnetic elements from the far-side of the Sun
\citep{2003SoPh..212..165S}. 

\end{enumerate}
A full discussion of these results along with their historical development can be found in the 
review by \cite{2005LRSP....2....5S}.

Some short term studies have used magnetic flux transport models to predict the nature of the polar 
magnetic field. These studies assume that the polar fields are solely due to the passive transport of magnetic 
flux from low to high
latitudes on the Sun. An example is  \cite{2001SoPh..201...57D,2002SoPh..211..103D} where the authors
consider a detailed study of high latitude magnetic plumes. 
These plumes are produced from poleward surges of magnetic flux that originate from activity complexes 
\citep{1983ApJ...265.1056G}. 
While a good agreement was found between observations and models, the authors did find some small discrepancies. 
They attributed these to small bipole emergences
outside the normal range of active latitudes considered in flux transport models. This indicates that the field at high 
latitudes is not solely the result of magnetic fields transported from low latitudes and a local dynamo action may 
play a role.
To date the only surface flux transport models to include such emergences are those of \citet{2001ApJ...547..475S} and \citet{2000SoPh..195..247W}.

In an additional study, \cite{2002SoPh..211...83D} and \cite{2003SoPh..214...23D} investigated in detail the reversal times of the polar magnetic fields, comparing the 
reversal times deduced from KP normal component magnetograms with those found in synoptic flux transport 
simulations.  To test whether magnetic observations or simulations gave the best estimate they deduced the 
locations of PILs at high latitudes from H$\alpha$ filament data. They then compared these to (i) the observed 
PIL locations deduced from Kitt Peak data and (ii) the simulated PILs from the synoptic flux transport equations. 
The study found that 
above 70\textdegree\ latitude the location of the PIL -- as given by the H$\alpha$ filaments -- was in fact better determined by the 
flux transport process rather than the direct magnetic field observations. This was due to the high level of uncertainty and error 
in measuring the field at high latitudes as a result of foreshortening and solar B angle effects \citep{2000SoPh..195..247W}. However, the simulation had to be run for over 20 rotations so that the high-latitude fields in 
the polar regions were solely the product of flux transport processes. This ensured that any systematic errors in the observations at high latitudes were removed.

A recent application of magnetic flux transport models has been to improve forecasts of the solar 10.7~cm (2.8~GHz) radio flux, using an empirical relation between this quantity and total ``sunspot'' and ``plage'' fields in the simulation \citep{2012SpWea..1002011H}. The 10.7~cm radio flux is widely used by the space weather community as a proxy for solar activity, with measurements dating back to 1947. For other aspects of space weather, flux transport models need to be coupled to coronal models: these applications will be discussed in Section~\ref{sec:app}.

\subsubsection{Multiple solar cycle applications of magnetic flux transport models}
\label{sec:appmftm2}

Agreement between observations and magnetic flux transport simulations is generally good over time-scales of
less than 11 years. But when magnetic flux transport simulations are run for multiple 11-year solar cycles, some 
inconsistencies are found. One of the first long term simulations was carried out by \cite{2002ApJ...577.1006S}
who simulated the photospheric field over 32 cycles from the Maunder minimum (1649) to 2002. With detailed
observations of magnetic bipoles available only for the last few solar cycles, the authors used 
synthetic data, scaling the activity level to the observed sunspot number. They also assumed that the flux transport 
parameters remained the same from one cycle to the next, and used a meridional flow profile that peaked at mid-latitudes 
with a value of 14~\ms. Since polar field production varies with the amount of emerging flux, the authors found that if the high latitude
field is solely described by the passive advection of magnetic flux from the active latitudes, 
then reversals of the polar field within each cycle may be lost. This is 
especially true if a series of weak cycles follows stronger ones. Although reversals were lost, they did return in later cycles, but the reversal in Cycle~23 did not match the observed time \citep[see Figure~1 of][]{2002ApJ...577.1006S}. To ensure that 
reversals occurred in 
every cycle, \cite{2002ApJ...577.1006S} introduced a new exponential decay term for the radial field in Equation \eqref{eq:ft}. This acts to reduce the strength
of the large scale field over long periods of time and prevents excess polar fields from building up. They found that
a decay time of 5\,--\,10~yr is optimal in allowing polar field reversals to occur from one cycle to the next.
Later, \citet{2006AA...446..307B} also found 
that a decay term was required in long term simulations to maintain the reversal of the polar field.  They formulated the 
decay term in terms of 3D diffusion of the magnetic field constrained on the 2D solar surface.

While \cite{2002ApJ...577.1006S} and \citet{2006AA...446..307B} introduced a new physical term into the 2D flux transport model, \cite{2002ApJ...577L..53W} considered a different approach. Rather than keeping the advection due to meridional flow constant from one cycle to the next, they varied the strength of meridional flow such that stronger cycles were given a faster flow. If a factor of two change occurs between strong and weak cycles, then reversals of the polar field may be maintained \citep[see Figure~2 of][]{2002ApJ...577L..53W}. This is because faster meridional flow leads to less flux canceling across the equator, and less net flux transported poleward. If correct, this suggests that in stronger cycles the magnetic fluxes in each hemisphere are more effectively isolated from one another. To complicate matters further, \cite{2010ApJ...719..264C} put forward a third possibility for maintaining polar field reversals. This was based around an observed cycle-to-cycle variation in the tilt angle of sunspot groups \citep{2010AA...518A...7D}. If there is an anti-correlation between cycle strength and tilt angles, where stronger cycles are assumed to have smaller tilt angles, then neither variable rates of meridional flow, or extra decay terms are required to maintain polar field reversals from one cycle to the next. To show this, \cite{2010ApJ...719..264C} consider a simulation extending from 1913\,--\,1986 and show that with such a tilt angle variation, the polar field reversal may be maintained. Although a decreased tilt angle alone was sufficient to maintain polar field reversals in this study, a longer term study by \cite{2011AA...528A..83J}  found that the radial decay term had to be re-introduced when carrying out simulations from 1700 to present \citep{2011AA...528A..82J}. The authors attributed this to inaccuracies in the input data of observed activity levels. However, it is not clear that observed tilt angles in the recent Cycle~23 were sufficiently reduced for this alone to explain the low polar fields in 2008 \citep{2008SoPh..252...19S}.

This discussion indicates that multiple combinations of model
parameters may produce qualitatively the same result. This is
illustrated by the study of \cite{2008SoPh..252...19S} who consider
the origin of the  decreased axial dipole moment found in 2008
compared to 1997. They simulated the global field from 1997\,--\,2008
and deduced that their previously included decay term was insufficient
by itself to account for the lower dipole moment. Instead, they found
that to reproduce the observations a steeper meridional flow gradient
is required at the equator. This steeper gradient effectively isolates
the hemispheres, as in  \cite{2002ApJ...577L..53W}. However, in
contrast to \cite{2002ApJ...577L..53W}, in their study
\cite{2008SoPh..252...19S} did not have to increase the meridional
flow rate, which was maintained at around 10~\ms. In a similar study,
\cite{2011SSRv..tmp..136J}  illustrated two additional ways of
reproducing the lower polar field strengths in 2008, namely (i) a 28\%
decrease in bipole tilt angles, or (ii) an increase in the meridional
flow rate from 11~\ms to 17~\ms. They, however, found that the first case
delayed the reversal time by 1.5~yr, so was inconsistent with
observations. Unfortunately, all of these possible solutions are
within current observational limits, so none can be ruled out.

\subsubsection{Variations in the meridional flow profile}
\label{sec:appmftm3}

While \cite{2002ApJ...577L..53W} and \cite{2008SoPh..252...19S} introduce overall variations of the
meridional flow profile and rate, they maintain a basic poleward flow profile in each hemisphere. Recent studies have considered more significant (and controversial) changes to the meridional flow, motivated by helioseismic observations. Two broad types of variation have been considered: a counter-cell near the poles, and an inflow towards activity regions.

\epubtkImage{mackay_star1.png}{%
\begin{figure}[htbp]
\centering\includegraphics[scale=0.8]{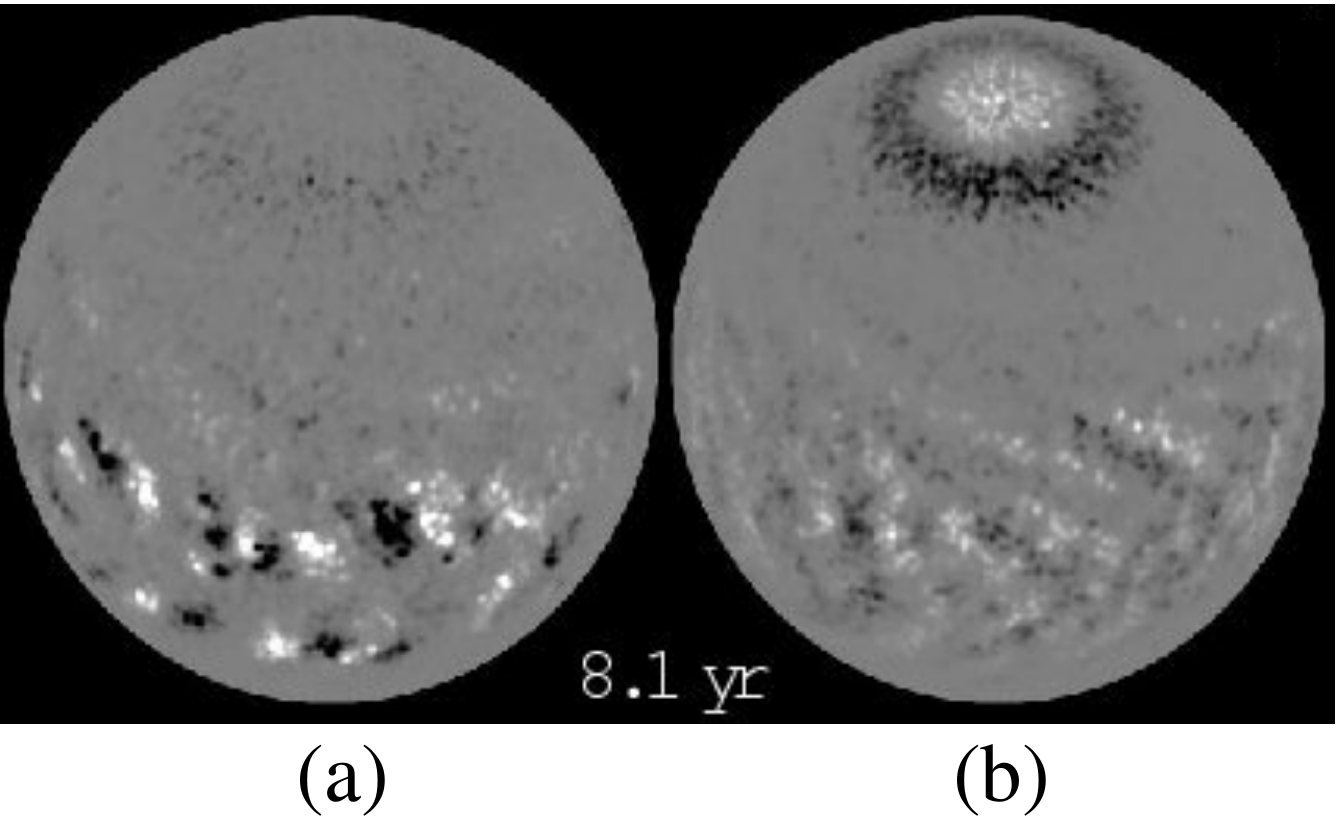}
\caption{Comparison of (a) solar and (b) stellar magnetic field
  configurations from \citet{2001ApJ...551.1099S}. For the stellar
  system all flux transport parameters are held fixed to solar values
  and only the emergence rate is increased to 30 times solar
  values. For the case of the Sun the magnetic fields saturate at
  \textpm~70~\Mxcm and for the star \textpm~700~\Mxcm. Image
  reproduced by permission from~\cite{2001ApJ...551.1099S}, copyright
  by AAS.}
\label{fig:star1}
\end{figure}}

\cite{2009ApJ...693L..96J} considered the possibility of the
existence of a counter-cell of reversed meridional flow near the poles. This change was motivated directly by observations showing that in Cycles~21 and 22 the polar field was strongly
peaked at the poles, while in Cycle~23 it peaked at a latitude of 75\textdegree\ and then reduced by nearly 50\% closer
to the pole \citep{2008ASPC..383...49R}. While the observed structure in Cycles~21 and 22 was consistent with a 
meridional flow profile that either extended all
the way to the poles, or switched off at around  75\textdegree, the Cycle~23 structure was not. To investigate what form of meridional flow could reproduce this, \cite{2009ApJ...693L..96J} introduced a counter-cell of meridional flow 
beyond 70\textdegree\ latitude
and showed that the observed distribution of magnetic flux in Cycle~23 could be achieved
with a counter-cell rate of 2\,--\,5~\ms. The authors showed that the counter-cell need not be constant and that
similar results could be obtained if it existed only in the declining phase. To date existence of such a counter-cell
has not been verified beyond doubt even though some results from helioseismology support its existence
\citep{2002ApJ...570..855H,2006ApJ...638..576G}. Also, due to the difference in profile of polar magnetic fields 
in Cycles~21 and 22 it is unlikely that such a counter-cell existed in these cycles.

The second type of flow perturbation suggested by observations are activity-dependent inflows towards the central latitude of emergence of the butterfly diagram \citep{2008SoPh..251..241G,2008SoPh..252..235G}. At least in part, this modulation of the axisymmetric meridional flow component results from the cumulative effect of the sub-surface horizontal flows converging towards active regions \citep{2004ApJ...613.1253H}. In an initial study, \citet{2006ESASP.624E..12D} simulated the effect of (non-axisymmetric) inflows towards active regions by adding an advection term to the flux transport model, its velocity proportional to the horizontal gradient of radial magnetic field. They found that if surface speeds exceed 10~\ms then an unrealistic ``clumping'' of magnetic flux that is not observed occurs. More recently, \citet{2010ApJ...717..597J} have considered the effect of an axisymmetric flow perturbation towards active latitudes of speed 3\,--\,5~\ms, similar to perturbations observed during Cycle~23. The authors consider how such a flow effects the
polar field distribution though simulated solar cycles. The main effect is to decrease
the separation of magnetic bipoles, subsequently leading to more cancellation in each hemisphere and less net flux 
transport poleward. This leads to an 18\% decrease in the
polar field strength compared to simulations without it. While this is a significant decrease, the authors note that it cannot (alone) account for the weak polar fields observed at the end of Cycle~23 as those fields 
decreased by more than a factor of 2. However, in conjunction with a variable meridional flow from one cycle to the next
\citep{2002ApJ...577L..53W} or the steepening gradient of the flow \citep{2008SoPh..252...19S}, the inflow may have a 
significant effect. \citet{2010ApJ...720.1030C} show that one can incorporate such axisymmetric flow perturbations by setting the speed at a given time proportional to the latitudinal gradient of longitude-averaged $B_r$. Not only does this generate appropriate inflows, but these can explain the observed solar cycle variation of the $P_2^1$ Legendre component of meridional flow \citep{2010Sci...327.1350H} without changing the background flow speed.

\subsubsection{Stellar applications of magnetic flux transport models}
\label{sec:appmftm4}

Due to advances in measuring stellar magnetic fields, magnetic flux transport models have 
recently been applied in the stellar context. One of their main applications has been to consider how 
polar or high latitude intense field regions or stellar spots may arise in rapidly rotating stellar systems
\citep{2001ASPC..223.1302S}. Initial studies \citep{2001ApJ...551.1099S,2004MNRAS.354..737M} used as a 
starting point parameters from solar magnetic flux transport simulations. These parameters were then varied 
to reproduce the key observational properties of the radial magnetic fields on stars as deduced through ZDI 
measurements. In the paper of \cite{2001ApJ...551.1099S} the authors consider a very active cool star of 
period 6~days. A key feature of their simulations is that they fix all parameters to values determined for 
the Sun and only vary the emergence rate to be 30 times solar values. They show that even if solar emergence 
latitudes are maintained, then the flux transport effects of meridional flow and surface diffusion are 
sufficient to transport enough flux to the poles to produce a polar spot. In Figure~\ref{fig:star1}, a 
comparison of typical solar (Figure~\ref{fig:star1}a) and stellar (Figure~\ref{fig:star1}b) magnetic 
field  configurations from Figure~2 of \cite{2001ApJ...551.1099S} can be seen. For the case of the Sun the 
magnetic fields saturate at \textpm~70~\Mxcm and for the star \textpm~700~\Mxcm. 
Both images show the polar region from a latitude of 40\textdegree. A clear difference can be seen, where for 
the Sun the pole has a weak unipolar field. For the rapidly rotating star there is a unipolar spot with
a ring of strong opposite polarity flux around it. The existence of this ring is partly due to the 
non-linear 
surface diffusion used in \cite{2001ApJ...551.1099S} which causes intense field regions to diffuse
more slowly (see Section~\ref{sec:extmftm}).

In contrast to the unipolar  poles modeled by \cite{2001ApJ...551.1099S}, some ZDI observations
of rapidly rotating stars show intermingling of opposite polarities at high latitudes,
where intense fields of both polarities lie at the same latitude and are not nested. An example of this
can be seen in AB~Dor (Figure~\ref{fig:mackay1}b) which has a rotation period of 1/2~day. 
\cite{2004MNRAS.354..737M} showed that in order to produce strong 
intermingled polarities within the polar regions more significant changes are required relative to the solar
flux transport model. These include increasing the emergence latitude of new bipoles from 40\textdegree\ to 
70\textdegree\ and increasing the rate of meridional flow from 11~\ms to 100~\ms. Increased emergence latitudes are consistent with an enhanced
Coriolis force deflecting more flux poleward as it travels through the convection zone 
\citep{1992AA...264L..13S}. However, at the present time, the predicted enhanced meridional flow is still below
the level of detection. While the simulations of \cite{2001ApJ...551.1099S} and \cite{2004MNRAS.354..737M}
have successfully reproduced key features in the magnetic field distributions of rapidly rotating stars, 
as yet they do not directly simulate the observations as has been done for the Sun. Presently, there is 
insufficient input data on the emergence latitudes of new bipoles in order to carry out such simulations.
In the study of \cite{2007AA...464.1049I} the authors used magnetic flux transport simulations to
estimate the lifetime of starspots as they are transported across the surface of a star. The authors show
that many factors may effect the star spot lifetime, such as the latitude of emergence and
differential rotation rate. In particular the authors show that for rapidly rotating stars the
lifetime of spots may be 1\,--\,2~months, however the lifetime is lower for stars with strong differential
rotation (AB~Dor) compared to those with weak differential rotation (HR~1099).

In recent years magnetic flux transport simulations have been combined with models for the generation
and transport of magnetic flux in the stellar interior \citep{2006MNRAS.369.1703H,2011AA...528A.135I} to
predict interior properties on other stars. The first study to link the pre- and post-eruptive
transport properties was carried out by \cite{2006MNRAS.369.1703H}. In this study, the 
authors quantified what effect the enhanced meridional flow predicted by \cite{2004MNRAS.354..737M}
would have on the dynamics and displacement of magnetic flux tubes as they rise though the convective zone 
\citep{1986AA...166..291M,1994AA...281L..69S}. The authors found that the enhanced meridional flow
leads to a non-linear displacement of the flux tubes, where bipoles were displaced to emerge
at higher latitudes. This was consistent with the required higher latitude of emergence to
produce the desired intermingling in the poles. However, in a more detailed model, 
\cite{2011AA...528A.135I} combine a  thin-layer $\alpha$-$\Omega$ dynamo model with a convective 
zone transport model and surface flux transport model to provide a complete description of the evolution of 
magnetic field on stars. Through this combined model the authors show that for rapid rotators 
(period $\sim$~2~days), due to the effect of the Coriolis force, the surface signature of the emergence and 
subsequent transport may not be a clear signal of the dynamo action that created it. This has important
consequences for both the observation and interpretation of magnetic fields on other stars.

\newpage

\section{Coronal Magnetic Field Models}
\label{sec:cf}

The magnetic fields observed in the solar photosphere extend out into the corona, where
they structure the plasma, store free magnetic energy and produce a wide variety of
phenomena. While the distribution and strength of magnetic fields are routinely measured 
in the photosphere, the same is not true for the corona, where the low densities mean that 
such measurements are very rare \citep{2009SSRv..144..413C}. To understand the nature of 
coronal magnetic fields, theoretical models that use the photospheric 
observations as a lower boundary condition are required.

In this section, we survey the variety of techniques that have been developed to model the 
coronal magnetic field based on the input of photospheric magnetic fields. We limit our discussion 
to global models in spherical geometry, and do not consider, for example, models for the magnetic 
structure of single active regions. In future revisions of this review, we will extend our discussion to 
include observations of both coronal and prominence magnetic fields in order to determine  the validity 
of these models.  Applications of the global models are the subject of Section~\ref{sec:app}.

A difficulty arising with any such attempt at a global model based on \emph{observations} of the photospheric magnetic field, is that we can only observe one side of the Sun at a single time (except for some limited observations by STEREO), but need data for all longitudes. This problem may be addressed either by compiling time series of full-disk observations into a synoptic observed magnetogram, or by assimilating individual active regions into a time-dependent surface flux transport model (Section~\ref{sec:mftm}). In this section, we will assume that the photospheric distribution of $B_r(R_{\odot},\theta,\phi)$ at a given time has already been derived by one of these techniques, and consider how to construct the magnetic field of the corona.

Virtually all global models to date produce \emph{equilibria}. This is reasonable since the coronal magnetic evolution on a global scale is essentially quasi-static. The magnetic field evolves through a continuous sequence of equilibria because the Alfv\'{e}n speed in the corona (the speed at which magnetic disturbances propagate) is of the order of 1000~\kms, much faster than the large-scale surface motions that drive the evolution (1\,--\,3~\kms). To model this quasi-static evolution, nearly all models produce a series of independent ``single-time'' extrapolations from the photospheric magnetic fields at discrete time intervals, without regard to the previous magnetic connectivity that existed in the corona. The exception is the flux transport magneto-frictional model (Section~\ref{sec:nlfff}) which produces a continuous sequence of equilibria. In this model, the equilibrium assumption is violated locally for brief periods during dynamical events such as CMEs or flares. Individual CME events are becoming practical to simulate in fully time-dependent MHD codes, but equilibrium models are still the only feasible way to understand when and where they are likely to occur.

We divide the techniques into four categories. The first three are potential field source surface models (Section~\ref{sec:pfss}), force-free field models (Section~\ref{sec:nlfff}), and magnetohydrostatic models (Section~\ref{sec:mhs}). All of these models output only (or primarily) the magnetic field. The fourth category are full MHD models (Section~\ref{sec:mhdfull}), which aim to self-consistently describe both the magnetic field and other plasma properties.

\subsection{Potential field source surface models}
\label{sec:pfss}

The most straightforward and therefore most commonly used technique for modeling the global coronal magnetic field is the so-called Potential Field Source Surface (PFSS) model \citep{1969SoPh....6..442S,1969SoPh....9..131A}.  We note that an implementation of the PFSS model by \citet{2002JGRA..107.1154L}, using photospheric magnetogram data from Wilcox Solar Observatory, is available to run ``on demand'' at \url{http://ccmc.gsfc.nasa.gov/}. The key assumption of this model is that there is zero electric current in the corona. The magnetic field is computed on a domain $R_{\odot}\leq r \leq R_{\mathrm{ss}}$, between the photosphere, $R_{\odot}$, where the radial magnetic field distribution\epubtkFootnote{Originally, it was more common to match the line-of-sight magnetic field rather than the radial component, since this was thought to better utilise the observations \citep{1969SoPh....9..131A}. But following \citet{1992ApJ...392..310W} matching the radial component is now considered more appropriate, because the photospheric magnetic field appears to be approximately radial.} is specified and an outer ``source surface'', $R_{\mathrm{ss}}$, where the boundary conditions $B_\theta=B_\phi=0$ are applied. The idea of the source surface is to model the effect of the solar wind outflow, which distorts the magnetic field away from a current-free configuration above approximately $2\,R_{\odot}$ (once the magnetic field strength has fallen off sufficiently). Distortion away from a current-free field implies the presence of electric currents at larger radii: a self-consistent treatment of the solar wind requires full MHD description (Section~\ref{sec:mhd}), but an artificial ``source surface'' boundary approximates the effect on the magnetic field in the lower corona. The PFSS model has been used to study a wide variety of phenomena ranging from open flux (see Section~\ref{sec:open}), coronal holes (Section~\ref{sec:ch}), and coronal null points \citep{2009ApJ...704.1021C} to magnetic fields in the coronae of other stars \citep{2002MNRAS.333..339J}.  

The requirement of vanishing electric current density in the PFSS model means that $\nabla\times\mathbf{B}=0$, so that
\begin{equation}
\mathbf{B} = -\nabla\Psi,
\end{equation}
where $\Psi$ is a scalar potential. The solenoidal constraint $\nabla\cdot\mathbf{B}=0$ then implies that $\Psi$ satisfies Laplace's equation
\begin{equation}
\nabla^2\Psi = 0,
\end{equation}
with Neumann boundary conditions
\begin{equation}
\left.\frac{\partial\Psi}{\partial r}\right|_{r=R_{\odot}} = -B_r(R_{\odot},\theta,\phi), \qquad \left.\frac{\partial\Psi}{\partial \theta}\right|_{r=R_{\mathrm{ss}}} = \left.\frac{\partial\Psi}{\partial \phi}\right|_{r=R_{\mathrm{ss}}}=0.
\end{equation}
Thus, the problem reduces to solving a single scalar PDE. It may readily be shown that, for fixed boundary conditions, the potential field is unique and, moreover, that it is the magnetic field with the lowest energy ($\int_VB^2/(2\mu_0)dV$) for these boundary conditions.

Solutions of Laplace's equation in spherical coordinates are well known \citep{1962clel.book.....J}. Separation of variables leads to
\begin{equation}
\Psi(r,\theta,\phi) = \sum_{l=0}^\infty\sum_{m=-l}^{l} \left[ f_{lm}r^l + 
g_{lm}r^{-(l+1)} \right] P_l^m(\cos\theta)\, e^{im\phi},
\end{equation}
where the boundary condition at $R_{\mathrm{ss}}$ implies that $g_{lm} = -f_{lm}R_{\mathrm{ss}}^{2l+1}$ for 
each pair $l$, $m$, and $P_l^m(\cos \theta)$ are the 
associated Legendre polynomials. The coefficients $f_{lm}$ are fixed by the photospheric $B_r(R_{\odot},\theta,\phi)$ distribution to be

\begin{equation}
f_{lm} = - \Big(l R_{\odot}^{l-1} + (l+1)R_{\odot}^{-l-2}R_{\mathrm{ss}}^{2l+1} \Big)^{-1}b_{lm},
\end{equation}
where $b_{lm}$ are the spherical harmonic coefficients of this $B_r(R_{\odot},\theta,\phi)$ distribution.
The solution for the three field components may then be written as
\begin{eqnarray}
B_r(r,\theta,\phi) &=& \sum_{l=0}^\infty\sum_{m=-l}^l b_{lm} c_l(r)P_l^m(\cos\theta)\,e^{im\phi},\\
B_\theta(r,\theta,\phi) &=&  -\sum_{l=0}^\infty\sum_{m=-l}^l b_{lm} d_l(r)\frac{d P_l^m(\cos\theta)}{d\theta}\,e^{im\phi},\\
B_\phi(r,\theta,\phi) &=&  -\sum_{l=0}^\infty\sum_{m=-l}^l \frac{im}{\sin\theta}\,b_{lm}d_l(r)P_l^m(\cos\theta)\,e^{im\phi},
\end{eqnarray}
where
\begin{eqnarray}
c_l(r) &=& \left(\frac{r}{R_{\odot}} \right)^{-l-2}\left[\frac{l + 1 + l(r/R_{\mathrm{ss}})^{2l+1}}{l + 1 + l(R_{\odot}/R_{\mathrm{ss}})^{2l+1}} \right],\\
d_l(r) &=& \left(\frac{r}{R_{\odot}} \right)^{-l-2}\left[\frac{1 - (r/R_{\mathrm{ss}})^{2l+1}}{l + 1 + l(R_{\odot}/R_{\mathrm{ss}})^{2l+1}} \right].
\end{eqnarray}
In practice, only a finite number of harmonics $l=1,\ldots,L_{\mathrm{max}}$ are included in the
construction of the field, depending on the resolution of the input photospheric $B_r$ distribution. Notice that the higher the mode number $l$, the faster the mode falls off with radial distance. This implies that the magnetic field at larger $r$ is dominated by the low order harmonics. So calculations of the Sun's open flux (Section~\ref{sec:open}) with the PFSS model do not require large $L_{\mathrm{max}}$. An example PFSS extrapolation is shown in Figure~\ref{fig:mackay3}a.

\epubtkImage{schatten.png}{%
\begin{figure}[htbp]
\centering\includegraphics[width=\textwidth]{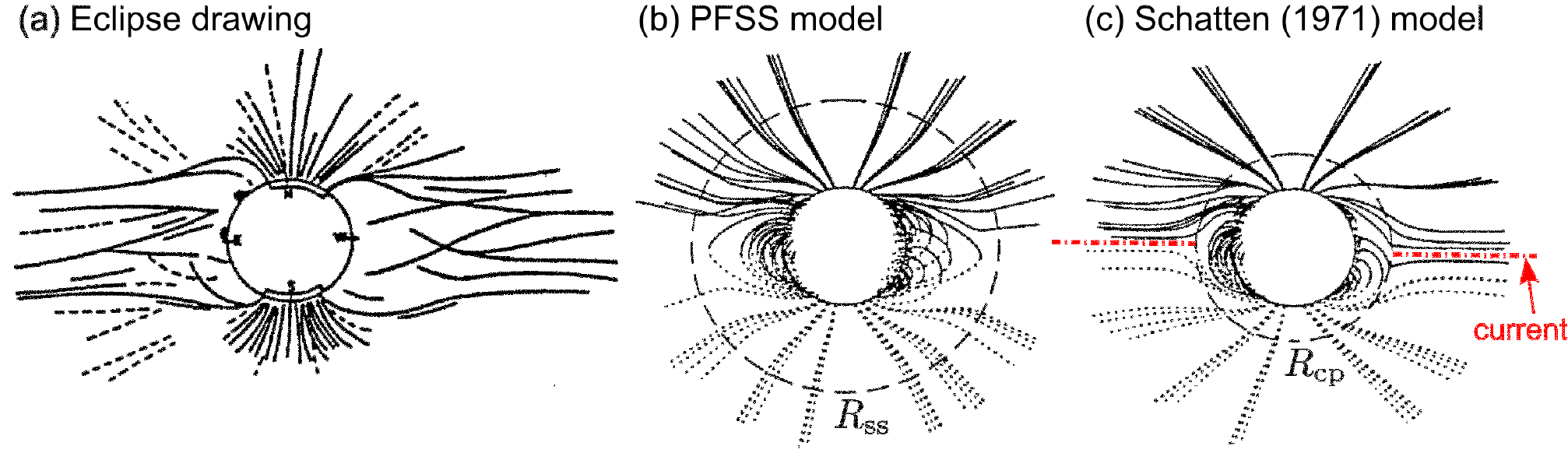}
\caption{Comparison between (a) a drawing of the real eclipse corona,
  (b) the standard PFSS model, and (c) the \citet{1971CosEl...2..232S}
  current sheet model. The current sheet model better reproduces the
  shapes of polar plumes and of streamer axes. Image adapted from
  Figure~1 of \citet{1994SoPh..151...91Z}.}
\label{fig:pfss}
\end{figure}}

The PFSS model has a single free parameter: the radius $R_{\mathrm{ss}}$ of the source surface. Various studies have chosen this parameter by optimising agreement with observations of either white-light coronal images, X-ray coronal hole boundaries, or, by extrapolating with \emph{in situ} observations of the interplanetary magnetic field (IMF). A value of $R_{\mathrm{ss}}=2.5\,R_{\odot}$ is commonly used, following \citet{1983JGR....88.9910H}, but values as low as $R_{\mathrm{ss}}=1.3\,R_{\odot}$ have been suggested \citep{1977ApJ...215..636L}. Moreover, it appears that even with a single criterion, the optimal $R_{\mathrm{ss}}$ can vary over time as magnetic activity varies \citep{2011SoPh..269..367L}.

A significant limitation of the PFSS model is that the actual coronal magnetic field does not become purely radial within the radius where electric currents may safely be neglected. This is clearly seen in eclipse observations (e.g., Figure~\ref{fig:pfss}a), as illustrated by \citet{1971CosEl...2..232S}, and in the early MHD solution of \citet{1971SoPh...18..258P}. Typically, real polar plumes bend more equatorward than those in the PFSS model, while streamers should bend more equatorward at Solar Minimum and more poleward at Solar Maximum. 

\citet{1971CosEl...2..232S} showed that the PFSS solution could be improved by replacing the source surface boundary $R_{\mathrm{ss}}$ with an intermediate boundary at $R_{\mathrm{cp}}=1.6\,R_{\odot}$, and introducing electric currents in the region $r>R_{\mathrm{cp}}$. To avoid too strong a Lorentz force, these currents must be limited to regions of weak field, namely to sheets between regions of $B_r>0$ and $B_r<0$. These current sheets support a more realistic non-potential magnetic field in $r>R_{\mathrm{cp}}$. The computational procedure is as follows:
\begin{enumerate}
\item Calculate $\mathbf{B}$ in the inner region $r\le R_{\mathrm{cp}}$ from the observed photospheric $B_r$, assuming an ``exterior'' solution (vanishing as $r\rightarrow\infty$).
\item Re-orientate $\mathbf{B}(R_{\mathrm{cp}},\theta,\phi)$ to ensure that $B_r>0$ everywhere (this will temporarily violate $\nabla\cdot\mathbf{B}=0$ on the surface $r=R_{\mathrm{cp}}$).
\item Compute $\mathbf{B}$ in the outer region $r>R_{\mathrm{cp}}$ using the exterior potential field solution, but matching all three components of $\mathbf{B}$ to those of the re-orientated inner solution on $R_{\mathrm{cp}}$. (These boundary conditions at $R_{\mathrm{cp}}$ are actually over-determined, so in practice the difference is minimised with a least-squares optimisation).
\item Restore the original orientation. This creates current sheets where $B_r=0$ in the outer region, but the magnetic stresses will balance across them (because changing the sign of all three components of $\mathbf{B}$ leaves the Maxwell stress tensor unchanged).
\end{enumerate}
This model generates better field structures, particularly at larger radii (Figure~\ref{fig:pfss}c), leading to its use in solar wind/space weather forecasting models such as the Wang--Sheeley--Arge (WSA) model \citep[][available at \url{http://ccmc.gsfc.nasa.gov/}]{2000JGR...10510465A}.
 The same field-reversal technique is used in the MHS-based Current-Sheet Source Surface model (Section~\ref{sec:mhs}), and was used also by \citet{1977SoPh...54..419Y}, who iterated to find a more realistic force-balance with plasma pressure and a steady solar wind flow.

By finding the surface where $B_r/|B|=0.97$ in their MHD model (Section~\ref{sec:mhdfull}), \citet{2006ApJ...653.1510R} locate the true ``source surface'' at $r\approx 13\,R_{\odot}$, comparable to the Alfv\'{e}n critical point where the kinetic energy density of the solar wind first exceeds the energy density of the magnetic field \citep{2010SoPh..266..379Z}. The Alfv\'{e}n critical points are known not to be spherically symmetric, and this is also found in the \citet{2006ApJ...653.1510R} model. In fact, a modified PFSS model allowing for a non-spherical source surface (though still at $2\mbox{\,--\,}3\,R_{\odot}$) was proposed by \citet{1978SoPh...60...83S}. For this case, before solving with the source surface boundary, the surface shape is first chosen as a surface of constant $|B|$ for the unbounded potential field solution.

\subsection{Nonlinear force-free field models}
\label{sec:nlfff}

While potential field solutions are straightforward to compute, the assumption of vanishing electric current density ($\mathbf{j}=0$) in the volume renders them unable to model magnetic structures requiring non-zero electric currents. Observations reveal such structures both in newly-emerged active regions, e.g., X-ray sigmoids, and outside active latitudes, including long-lived H$\alpha$ filament channels and coronal magnetic flux ropes. A significant limitation of the potential field is that it has the lowest energy compatible with given boundary conditions. Yet many important coronal phenomena derive their energy from that stored in the magnetic field; this includes large-scale eruptive events (flares and CMEs), but also small-scale dynamics thought to be responsible for heating the corona. We still lack a detailed understanding of how these events are initiated. Such an understanding cannot be gained from potential field models, owing to the lack of free magnetic energy available for release. The models in this section, based on the force-free assumption, allow for electric currents and, hence, free magnetic energy.

In the low corona of a star like the Sun, the magnetic pressure dominates over both the gas pressure and the kinetic energy density of plasma flows. Thus to first approximation an equilibrium magnetic field must have a vanishing Lorentz force,
\begin{equation}
\mathbf{j}\times\mathbf{B} = 0,
\end{equation}
where $\mathbf{j}=\nabla\times\mathbf{B}/\mu_0$ is the current density. Such a magnetic field is called \emph{force-free}. It follows that
\begin{equation}
\nabla\times\mathbf{B} = \alpha\mathbf{B}
\label{eqn:ffalpha}
\end{equation}
for some scalar function $\alpha(\mathbf{r})$ which is constant along magnetic field lines (this follows from $\nabla\cdot\mathbf{B}=0$). The function $\alpha(\mathbf{r})$ depends only on the local geometry of field lines, not on the field strength; it quantifies the magnetic ``twist'' (Figure~\ref{fig:twist}). The potential field $\mathbf{j}=0$ is recovered if $\alpha=0$ everywhere. 

\epubtkImage{ary_twist.png}{%
\begin{figure}[htbp]
\centering\includegraphics[width=0.6\textwidth]{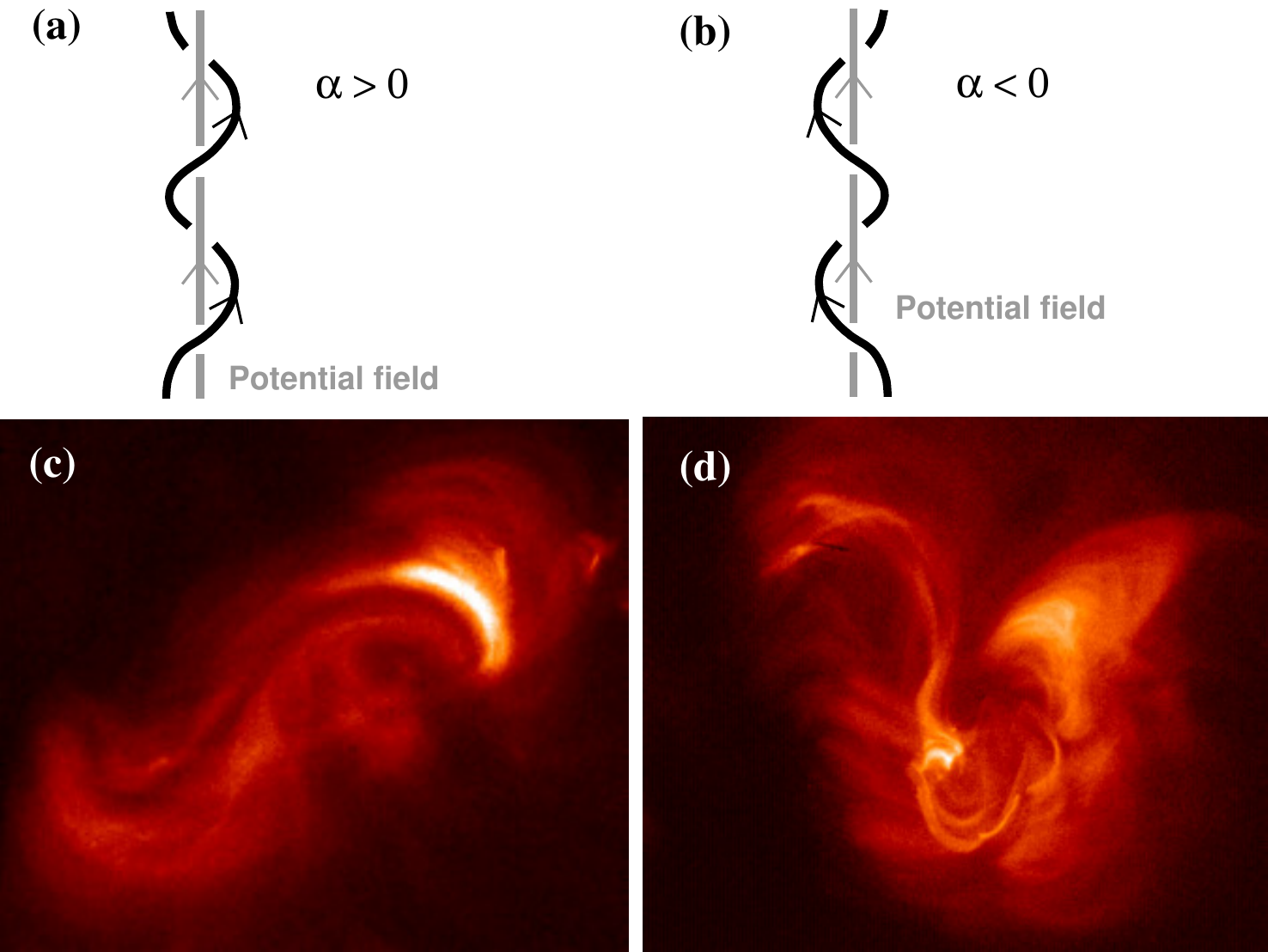}
\caption{Twisted magnetic fields. Panels (a) and (b) show the direction of twist for a force-free field line with (a) positive $\alpha$, and (b) negative $\alpha$, with respect to the potential field $\alpha=0$. Panels (c) and (d) show X-ray sigmoid structures with each sign of twist, observed in active regions with the Hinode X-ray Telescope (SAO, NASA, JAXA, NAOJ). These images were taken on 2007 February 16 and 2007 February 5, respectively.}
\label{fig:twist}
\end{figure}}

Unfortunately, for non-zero $\alpha$, the extrapolation of coronal force-free magnetic fields is not straightforward. If $\alpha$ is constant everywhere, then taking the curl of (\ref{eqn:ffalpha}) leads to the vector Helmholtz equation
\begin{equation}
(\nabla^2 + \alpha^2)\mathbf{B} = 0,
\end{equation}
which is linear and may be solved analytically in spherical harmonics
\citep{1989AuJPh..42..317D}. While some early attempts were made to
model the global corona using such \emph{linear} force-free fields
\citep{1973AA....27...95N,1974SoPh...36..345L}, their use is limited
for two main reasons. Firstly, the constant $\alpha$ in such a
solution scales as $L^{-1}$, where $L$ is the horizontal size of the
area under consideration. So for global solutions, only rather small
values of $\alpha$, close to potential, can be used. Secondly,
observations of twist in magnetic structures indicate that both the
sign and magnitude of $\alpha$ ought to vary significantly between
different regions on the Sun. In addition, a mathematical problem
arises if one attempts to apply the same boundary conditions as in the
potential field model (namely a given distribution of $B_r$ on
$r=R_{\odot}$, and $B_\theta=B_\phi=0$ on $r=R_{\mathrm{ss}}$). A
strictly force-free field satisfying $B_\theta=B_\phi=0$ and $B_r\neq
0$ on an outer boundary must have $\alpha=0$ on that boundary
\citep{1993SoPh..144..243A}, so that $\alpha=0$ on all open magnetic
field lines. Unless $\alpha=0$, this is incompatible with the linear
force-free field.%
\epubtkFootnote{\citet{1989AuJPh..42..317D} suggested instead to
  minimise the horizontal field on $r=R_{\mathrm{ss}}$ in a
  least-squares sense.}
From this, it is clear that the use of \emph{linear} force-free fields is restrictive and will not be discussed further. Instead, we will consider \emph{nonlinear} force-free fields where $\alpha$ is a function of position.

The problem of extrapolating nonlinear force-free fields from given photospheric data is mathematically challenging, with open questions about the existence and uniqueness of solutions. Nevertheless, several numerical techniques have been developed in recent years, though they have largely been applied to single active regions in Cartesian geometry \citep{2006SoPh..235..161S,2009ApJ...696.1780D}. Applications to global solutions in spherical geometry are in their infancy: we describe here the three main approaches tried.

\subsubsection{Optimisation method}

\citet{2007SoPh..240..227W} has developed a method for numerically computing nonlinear force-free fields in spherical geometry, based on the optimisation procedure of \citet{2000ApJ...540.1150W} and using the \emph{vector} magnetic field in the photosphere as input. The idea is to minimise the functional
\begin{equation}
L[\mathbf{B}] = \int_V\left(\frac{|(\nabla\times\mathbf{B})\times\mathbf{B}|^2}{B^2} + |\nabla\cdot\mathbf{B}|^2 \right)dV,
\label{eqn:lfunc}
\end{equation}
since if $L=0$ then $\mathbf{j}\times\mathbf{B}=0$ and $\nabla . \mathbf{B}=0$ everywhere in the volume $V$. Differentiating (\ref{eqn:lfunc}) with respect to $t$, \citet{2000ApJ...540.1150W} show that
\begin{equation}
\frac{1}{2}\frac{dL}{dt} = -\int_V\frac{\partial\mathbf{B}}{\partial t}\cdot\mathbf{F}\,dV - \oint_{\partial V}\frac{\partial\mathbf{B}}{\partial t}\cdot\mathbf{G}\,dS,
\end{equation}
where
\begin{eqnarray}
\mathbf{F} &=& \nabla\times(\boldsymbol{\Omega}\times\mathbf{B}) - \boldsymbol{\Omega}\times(\nabla\times\mathbf{B}) - \nabla(\boldsymbol{\Omega}\cdot\mathbf{B}) + \boldsymbol{\Omega}(\nabla\cdot\mathbf{B}) + \boldsymbol{\Omega}^2\mathbf{B},\\
\mathbf{G} &=& \mathbf{n}\times(\boldsymbol{\Omega}\times\mathbf{B}) - \mathbf{n}(\boldsymbol{\Omega}\cdot\mathbf{B}),\\
{\bf \Omega} &=& \frac{1}{B^2}\Big[(\nabla\times\mathbf{B})\times\mathbf{B} - (\nabla\cdot\mathbf{B})\mathbf{B} \Big].
\end{eqnarray}
Evolving the magnetic field so that
\begin{equation}
\frac{\partial\mathbf{B}}{\partial t} = \mu\mathbf{F}
\label{eqn:wiegel}
\end{equation}
for some $\mu>0$, and imposing $\partial\mathbf{B}/\partial t = 0$ on the boundary $\partial V$, will ensure that $L$ decreases during the evolution. To apply this procedure, \citet{2007SoPh..240..227W} takes an initial potential field extrapolation, and replaces $B_\theta$ and $B_\phi$ on $r=R_{\odot}$ with the measured horizontal magnetic field from an observed vector magnetogram. The 3D field is then iterated with Equation~(\ref{eqn:wiegel}) until $L$ is suitably small.

\citet{2007SoPh..240..227W} demonstrates that this method recovers a known analytical force-free field solution (Figure~\ref{fig:opt}), although the most accurate solution is obtained only if the boundary conditions at both the photosphere and upper source surface are matched to the analytical field. The method is yet to be applied to observational data on a global scale, largely due to the limitation that vector magnetogram data are required for the lower boundary input. Such data are not yet reliably measured on a global scale, though SDO or SOLIS should be a key step forward in obtaining this data. Another consequence is that even where they are measured there is the complication that the photospheric magnetic field is not force-free. Pre-processing techniques to mitigate these problems are under development \citep{2011AA...527A..30T}.

\epubtkImage{ary_opt.png}{%
\begin{figure}[htbp]
\centering\includegraphics[width=\textwidth]{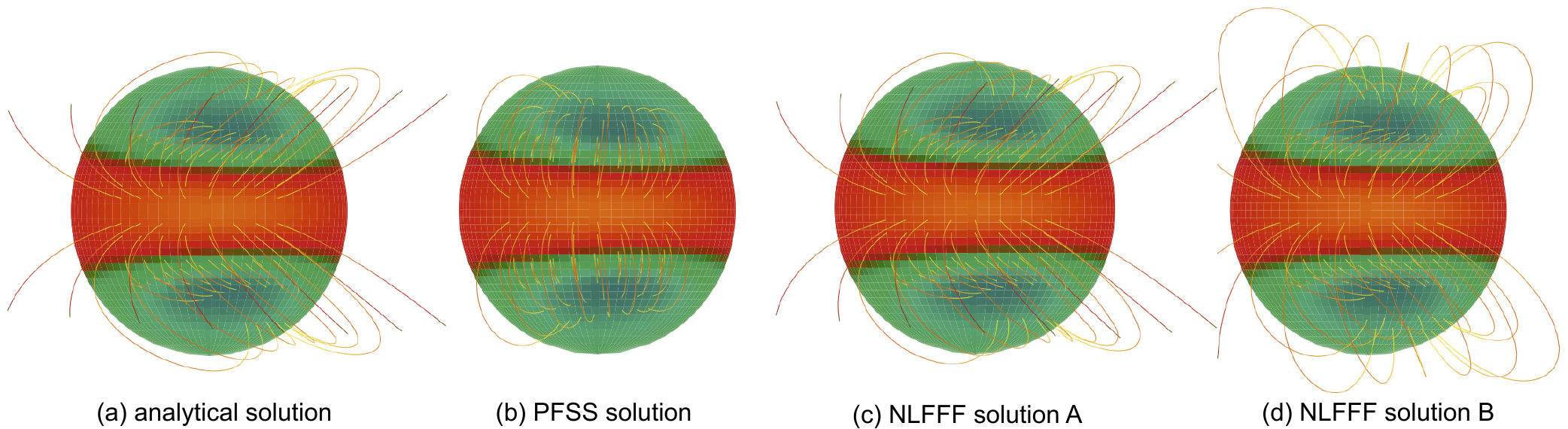}
\caption{Test of the optimisation method for nonlinear force-free fields with an analytical solution \citep[figure adapted from][]{2007SoPh..240..227W}. Panel (a) shows the analytical solution \citep{1990ApJ...352..343L} and (b) shows the PFSS extrapolation used to initialise the computation. Panel (c) shows the resulting NLFFF when the boundary conditions at both $r=R_{\odot}$ and $r=R_{\mathrm{ss}}$ are set to match the analytical solution, while panel (d) shows how the agreement is poorer if the analytical boundary conditions are applied only at the photosphere (the more realistic case for solar application).}
\label{fig:opt}
\end{figure}}

\subsubsection{Force-free electrodynamics method}

\citet{2011SoPh..269..351C} have recently proposed an alternative method to compute global nonlinear force-free field extrapolations, adapted from a technique applied to pulsar magnetospheres. It has the advantage that only the radial component $B_r$ is required as observational input at the photosphere. The computation is initialised with an arbitrary 3D magnetic field such as $\mathbf{B}=0$ or that of a dipole, and also with zero electric field $\mathbf{E}$. The fields $\mathbf{E}$ and $\mathbf{B}$ are then evolved through the equations of \emph{force-free electrodynamics},
\begin{eqnarray}
\frac{\partial\mathbf{B}}{\partial t} &=& -c\nabla\times\mathbf{E},\label{eqn:ffe1}\\
\frac{\partial\mathbf{E}}{\partial t} &=& c\nabla\times\mathbf{B} - 4\pi\mathbf{j},\label{eqn:ffe2}\\
\nabla\cdot\mathbf{B} &=& 0,\\
\rho_e\mathbf{E} + \frac{1}{c}\mathbf{j}\times\mathbf{B} &=& 0, \label{eqn:ffe3}
\end{eqnarray}
where $\rho_e=\nabla\cdot\mathbf{E}/(4\pi)$ is the electric charge density. The force-free condition (\ref{eqn:ffe3}) enables $\mathbf{j}$ to be eliminated from (\ref{eqn:ffe2}) since it follows that
\begin{equation}
\mathbf{j} = \frac{c}{4\pi}\nabla\cdot\mathbf{E}\frac{\mathbf{E}\times\mathbf{B}}{B^2} + \frac{c}{4\pi}\left(\frac{\mathbf{B}\cdot\nabla\times\mathbf{B} - \mathbf{E}\cdot\nabla\times\mathbf{E}}{B^2}\right)\mathbf{B}.
\end{equation}
Equations~(\ref{eqn:ffe1}) and (\ref{eqn:ffe2}) are integrated numerically with time-dependant $E_\theta$, $E_\phi$ imposed on the lower boundary $r=R_{\odot}$. These are chosen so that $B_r$ gradually evolves toward its required distribution. The photospheric driving injects electrodynamic waves into the corona, establishing a network of coronal electric currents. The outer boundary is chosen to be non-reflecting and perfectly absorbing, mimicking empty space. As the photosphere approaches the required distribution, the charge density and electric fields diminish, but the coronal currents remain. As $\mathbf{E}\rightarrow 0$ a force-free field is reached. 

The authors have tested their method with an observed synoptic magnetogram: an example force-free field produced is shown in Figure~\ref{fig:ffed}. They find that active region magnetic fields are reproduced quite rapidly, but convergence to the weaker fields in polar regions is much slower. Slow convergence is undesirable because numerical diffusion was found to erode the coronal currents before the photosphere reached the target configuration. However, the convergence rate could be improved by choosing the initial condition to be a dipole, approximating the average polar field in the magnetogram. This highlights an important feature of the method: \emph{non-uniqueness}. The force-free field produced is not defined solely by the photospheric boundary condition, but depends both on (i) the choice of initialisation and (ii) the path followed to reach the final state. \citet{2011SoPh..269..351C} suggest that one way to choose between possible solutions would be to incorporate measurements from vector magnetograms. The model described in the next section takes a different approach in treating the construction of the coronal magnetic field as an explicitly time-dependent problem.

\epubtkImage{ary_ffed.png}{%
\begin{figure}[htbp]
\centering\includegraphics[width=0.6\textwidth]{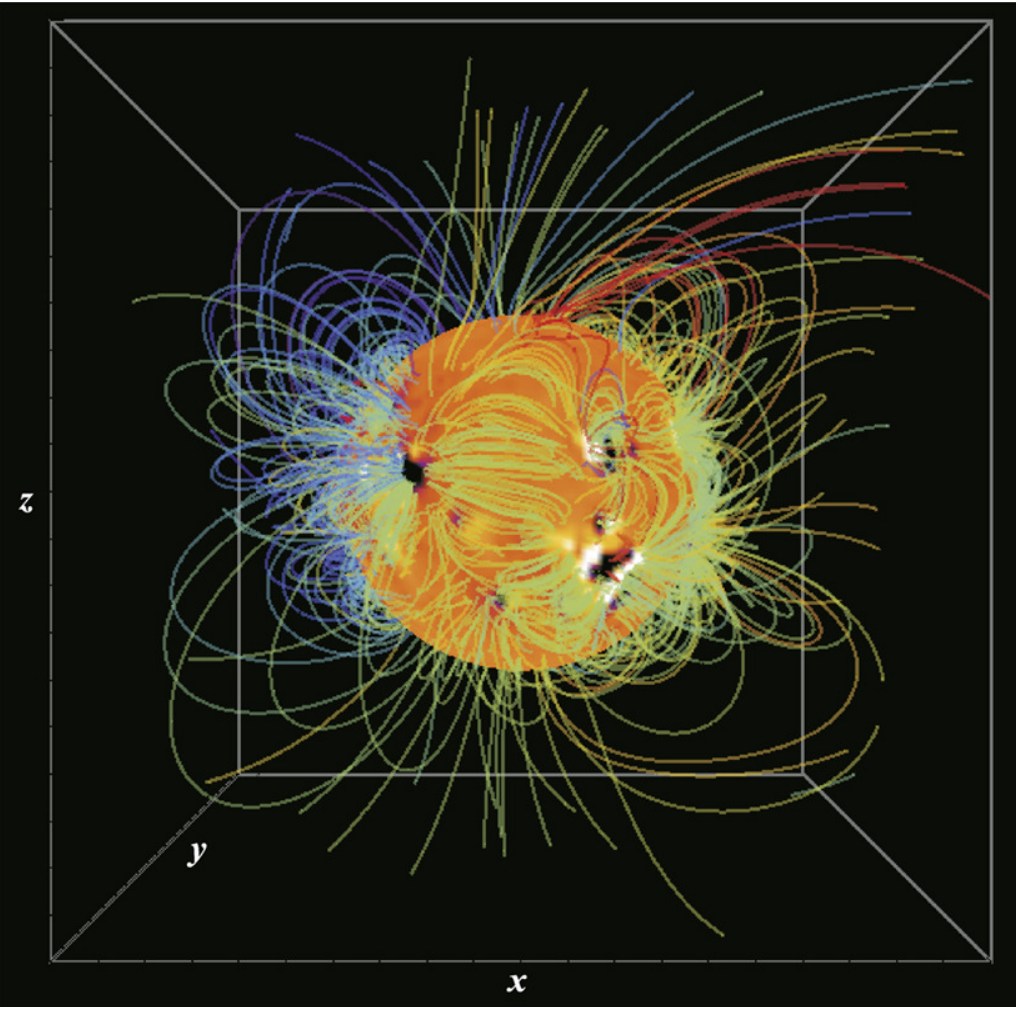}
\caption{A force-free magnetic field produced by the force-free
  electrodynamics method for Carrington rotation 2009. Field line
  colours show the twist parameter $\alpha$, coloured red where
  $\alpha>0$, blue where $\alpha<0$, and light green where
  $\alpha\approx 0$. Image reproduced by permission from
  \citet{2011SoPh..269..351C}, copyright by Springer.}
\label{fig:ffed}
\end{figure}}

\subsubsection{Flux transport and magneto-frictional method}
\label{sec:ftmf}

Recently, \citet{2000ApJ...539..983V} and \citet{2006ApJ...641..577M} have developed a new technique to study the long-term evolution of coronal magnetic fields, which has now been applied to model the global solar corona \citep{2007SoPh..245...87Y,2008SoPh..247..103Y}. The technique follows the build-up of free magnetic energy and electric currents in the corona by coupling together two distinct models. The first is a data driven surface flux transport model \citep{2007SoPh..245...87Y}. This uses observations of newly emerging magnetic bipoles to produce a continuous evolution of the observed photospheric magnetic flux over long periods of time. Coupled to this is a quasi-static coronal evolution model \citep{2006ApJ...641..577M,2008SoPh..247..103Y} which evolves the coronal magnetic field through a sequence of nonlinear force-free fields in response to the observed photospheric evolution and flux emergence. The model follows the long-term continuous build-up of free magnetic energy and electric currents in the corona. It differs significantly from the extrapolation approaches which retain no memory of magnetic flux or connectivity from one extrapolation to the next. 

The photospheric component of the model evolves the radial magnetic field $B_r$ on $r=R_{\odot}$ with a standard flux transport model (Section~\ref{sec:basicmftm}), except that newly emerging bipolar active regions are inserted not just in the photosphere but also in the 3D corona. These regions take an analytical form \citep{2007SoPh..245...87Y}, with parameters (location, size, flux, tilt angle) chosen to match observed active regions. An additional twist parameter allows the emerging 3D regions to be given a non-zero helicity: in principle this could be determined from vector magnetogram observations, but these are not yet routinely available.

The coronal part of the model evolves the large-scale \emph{mean} field \citep{2000ApJ...539..983V} according to the induction equation
\begin{equation}
\frac{\partial \mathbf{A}}{\partial t} = \mathbf{v}\times\mathbf{B} + {\cal E},
\end{equation}
where $\mathbf{B}=\nabla\times\mathbf{A}$ is the mean magnetic field (with the vector potential $\mathbf{A}$ in an appropriate gauge). The mean electromotive force ${\cal E}$ describes the net effect of unresolved small-scale fluctuations, for example, braiding and current sheets produced by interaction with convective flows in the photosphere. \citet{2006ApJ...641..577M} and \citet{2008SoPh..247..103Y} assume the form
\begin{equation}
{\cal E} = -\eta\mathbf{j},
\end{equation}
where
\begin{equation}
\eta = \eta_0\left(1 + 0.2\frac{|\mathbf{j}|}{B} \right)
\end{equation}
is an effective turbulent diffusivity. The first term is a uniform background value $\eta_0=45\mathrm{\ km^{2}\ s^{-1}}$ and the second term is an enhancement in regions of strong current density, introduced to limit the twist in helical flux ropes to about one turn, as observed in solar filaments.

Rather than solving the full MHD equations for the velocity $\mathbf{v}$, which is not computationally feasible over long timescales, the quasi-static evolution of the coronal magnetic field is approximated using magneto-frictional relaxation \citep{1986ApJ...309..383Y}, setting
\begin{equation}
\mathbf{v} = \frac{1}{\nu}\frac{\mathbf{j}\times\mathbf{B}}{B^2}.
\end{equation}
This artificial velocity causes the magnetic field to relax towards a force-free configuration: it may be shown that the total magnetic energy decreases monotonically until $\mathbf{j}\times\mathbf{B}=0$. Here, the magneto-frictional relaxation is applied concurrently with the photospheric driving so, in practice, a dynamical equilibrium between the two is reached. At the outer boundary $r=2.5\,R_{\odot}$, \citet{2006ApJ...641..577M} introduced an imposed radial outflow, rather than setting $B_\theta=B_\phi=0$ exactly. This allows magnetic flux ropes to be ejected after they lose equilibrium (simulating CMEs on the real Sun), and simulates the effect of the solar wind in opening up magnetic field lines radially above this height.

\epubtkImage{magnetofric.png}{%
\begin{figure}[htbp]
\centering\includegraphics[scale=0.6]{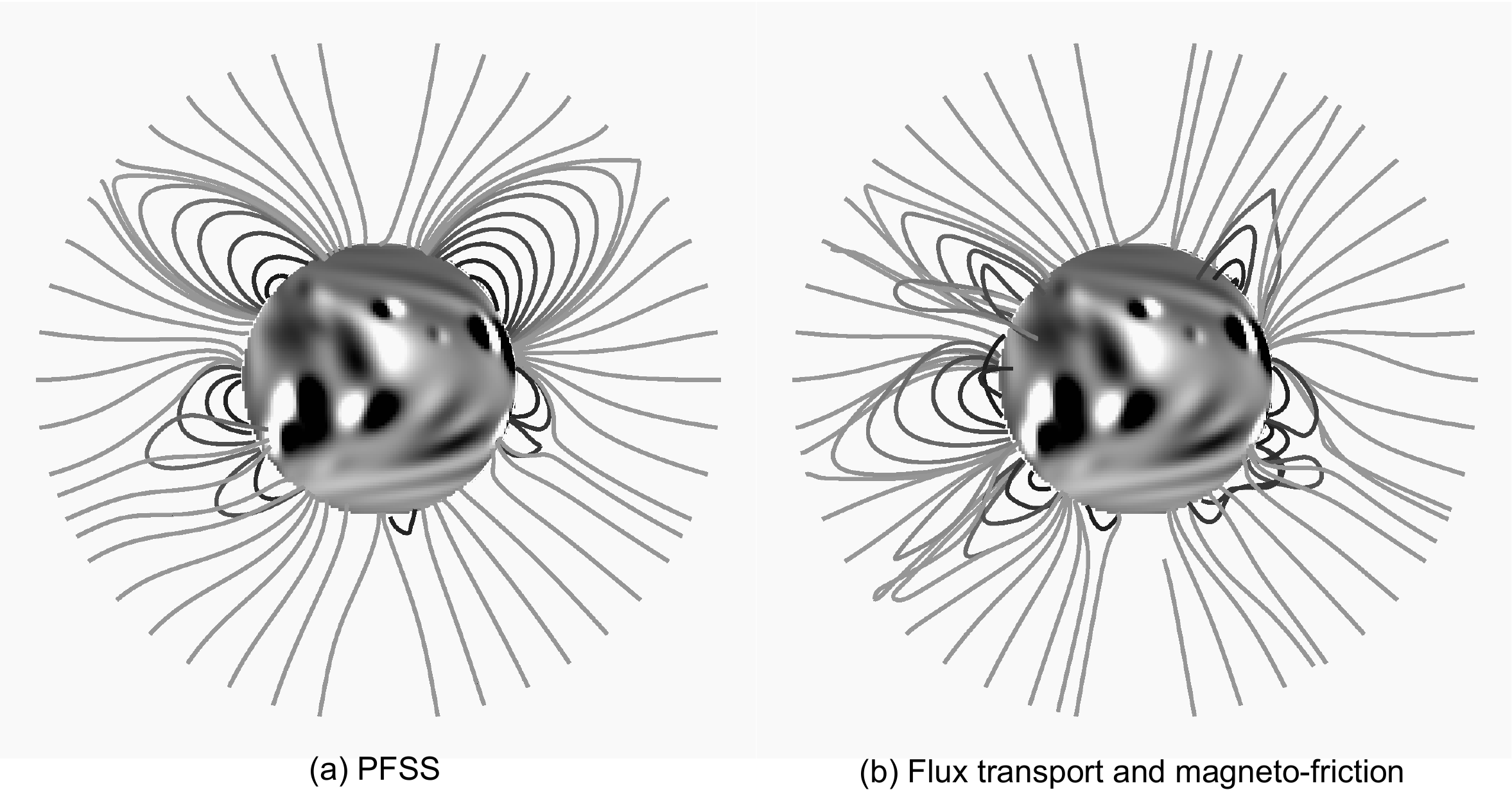}
\caption{(a) Example of a Potential Field Source Surface
  extrapolation. (b) Example of a non-potential coronal field produced
  by the global coronal evolution model of \cite{2006ApJ...641..577M}
  and \cite{2008SoPh..247..103Y}. The grey-scale image shows the
  radial field at the photosphere and the thin lines the coronal field
  lines. Image reproduced by permission
  from~\cite{2010JGRA..11509112Y}, copyright by AGU.}
\label{fig:mackay3}
\end{figure}}

Figure~\ref{fig:mackay3}b shows an example nonlinear force-free field using this model, seen after 100~days of evolution. Figure~\ref{fig:mackay3}a shows a PFSS extrapolation from the same photospheric $B_r$ distribution. The coronal field in the non-potential model is significantly different, comprising highly twisted flux ropes, slightly sheared coronal arcades, and near potential open field lines. Since this model can be run for extended periods without resetting the coronal field, it can be used to study long-term helicity transport across the solar surface from low to high latitudes \citep{2008ApJ...680L.165Y}. 

There are several parameters in the model. As in the PFSS model the location of the upper boundary $R_{\mathrm{ss}}$ is arbitrary, although it has less influence since the non-potential magnetic field strength falls off more slowly with radius than that of the potential field. In the magneto-frictional evolution, the turbulent diffusivity $\eta_0$ and friction coefficient $\nu$ are arbitrary, and must be calibrated by comparison with observed structures or timescales for flux ropes to form or lose equilibrium. Finally a 3D model is required for newly-emerging active regions: the simple analytical bipoles of the existing simulations could in fact be replaced with more detailed extrapolations from observed photospheric fields in active regions.

\subsection{Magnetohydrostatic models}
\label{sec:mhs}

Models in this category are based on analytical solutions to the full magnetohydrostatic (MHS) equations,
\begin{eqnarray}
\mathbf{j}\times\mathbf{B} - \nabla p - \rho\nabla\psi &=& 0,\\
\nabla\times\mathbf{B} &=& \mu_0\mathbf{j},\\
\nabla\cdot\mathbf{B} &=& 0,
\end{eqnarray}
where for the coronal application $\psi=-GM/r$ is taken to be the gravitational potential. Three-dimensional solutions that are general enough to accept arbitrary $B_r(R_{\odot},\theta,\phi)$ as input have been developed by \citet{1986ApJ...306..271B} and further by \citet{1995AA...301..628N}. The basic idea is to choose a particular functional form of $\mathbf{j}$, and use the freedom to choose $\rho$ and $p$ to obtain an analytical solution. Note that, while the distributions of $\rho$, $p$, and $\mathbf{B}$ are self-consistent in these solutions, the 3D forms of $\rho$ and $p$ are prescribed in the solution process. In particular, they cannot be constrained \emph{a priori} to satisfy any particular equation of state or energy equation. A realistic treatment of the thermodynamics of the plasma requires a full MHD model (Section~\ref{sec:mhdfull}).

The solutions that have been used to extrapolate photospheric data have a current density of the form
\begin{equation}
\mathbf{j} = \alpha\mathbf{B} + \xi(r)\nabla(\mathbf{r}\cdot\mathbf{B})\times\mathbf{e}_r,
\end{equation}
where
\begin{equation}
\xi(r) = \frac{1}{r^2} - \frac{1}{(r+a)^2}
\end{equation}
and $\alpha$, $a$ are constant parameters.
Thus, the current comprises a field-aligned part and a part perpendicular to gravity. For this form of $\mathbf{j}$, \citet{1995AA...301..628N} shows that the solution takes the form
\begin{eqnarray}
B_r &=& \sum_{l=1}^\infty\sum_{m=-l}^l\sum_{j=1}^2 A_{lm}^{(j)}\frac{l(l+1)}{r}u_l^{(j)}(r)P_l^m(\cos\theta)\,e^{im\phi},\\
B_\theta &=& \sum_{l=1}^\infty\sum_{m=-l}^l\sum_{j=1}^2 A_{lm}^{(j)} \left[\frac{1}{r}\frac{d(ru_l^{(j)}(r))}{dr}\frac{dP_l^m(\cos\theta)}{d\theta} + \frac{\mu_0\alpha mi}{\sin\theta}u_l^{(j)}(r)P_l^m(\cos\theta) \right]\,e^{im\phi},\\
B_\phi &=& \sum_{l=1}^\infty\sum_{m=-l}^l\sum_{j=1}^2 A_{lm}^{(j)} \left[\frac{im}{r\sin\theta}\frac{d(ru_l^{(j)}(r))}{dr}P_l^m(\cos\theta) - \mu_0\alpha u_l^{(j)}(r)\frac{dP_l^m(\cos\theta)}{d\theta}  \right]\,e^{im\phi},
\end{eqnarray}
where
\begin{eqnarray}
u_l^{(1)}(r) &=& \frac{\sqrt{r+a}}{r}J_{l+1/2}\Big(\alpha(r+a)\Big),\\
u_l^{(2)}(r) &=& \frac{\sqrt{r+a}}{r}N_{l+1/2}\Big(\alpha(r+a)\Big).
\end{eqnarray}
Here, $J_{l+1/2}$ and $N_{l+1/2}$ are Bessel functions of the first and second kinds, respectively. The coefficients $A_{lm}^{(j)}$ are determined from the boundary conditions, as in the PFSS model. The plasma pressure and density then take the forms
\begin{eqnarray}
p(r,\theta,\phi) &=& p_0(r) -\frac{\xi(r)}{2}(\mathbf{r}\cdot\mathbf{B})^2,\\
\rho(r,\theta,\phi) &=& \rho_0(r) + \frac{r^2}{GM}\left(\frac{1}{2}\frac{d\xi(r)}{dr}(\mathbf{r}\cdot\mathbf{B})^2 + r\xi(r)\mathbf{B}\cdot\nabla(\mathbf{r}\cdot\mathbf{B}) \right). 
\end{eqnarray}
The functions $p_0(r)$ and $\rho_0(r)$ describe a spherically symmetric background atmosphere satisfying $-\nabla p_0 - \rho_0\nabla\psi=0$, and may be freely chosen.

Solutions of this form have been applied to the coronal extrapolation problem by \citet{2000ApJ...538..932Z}, \citet{2001SoPh..198..279R}, and \citet{2008AA...481..827R}.%
\epubtkFootnote{The expression for $\mathbf{B}$ quoted by \citet{2000ApJ...538..932Z} contains a typographical error (in their Equation~2), as does that by \citet{2008AA...481..827R} (in their Equation~8).}
These solutions are of exterior type, i.e., with $|B|\rightarrow 0$ as $r\rightarrow\infty$. For $\alpha\neq 0$, the latter condition follows from the properties of both types of Bessel function. 

There are two free parameters in the solution that may be varied to best fit observations: $a$ and $\alpha$. Broadly speaking, the effect of increasing $a$ is to inflate/expand the magnetic field, while the effect of increasing $|\alpha|$ is to twist/shear the magnetic field. This is illustrated in Figure~\ref{fig:zhao2000}. Note that too large a value of $\alpha$ would lead to zeros of the Bessel functions falling within the computational domain, creating magnetic islands that are unphysical for the solar corona \citep{1995AA...301..628N}. The (unbounded) potential field solution $\mathbf{j}=0$ is recovered for $\alpha=a=0$, while if $a=0$ then $\mathbf{j}=\alpha\mathbf{B}$ and we recover the linear force-free field case. If $\alpha=0$ then the current is purely horizontal and the solution reduces to Case~III of \citet{1986ApJ...306..271B}. \citet{1995JGR...10019865G} applied this earlier solution to the Solar Minimum corona, although they showed that it was not possible to match both the density distribution in the corona and the photospheric magnetic field to observations. This problem is likely to be exacerbated at Solar Maximum. Similarly, \citet{2008AA...481..827R} found that the strongest density perturbation in this model appears in active regions in the low corona. Preventing the density from becoming negative can require an unrealistically large background density $\rho_0$ at these radii, particularly for large values of the parameter $a$.

\epubtkImage{ary_zhao2000.png}{%
\begin{figure}[htb]
\centering\includegraphics[width=0.8\textwidth]{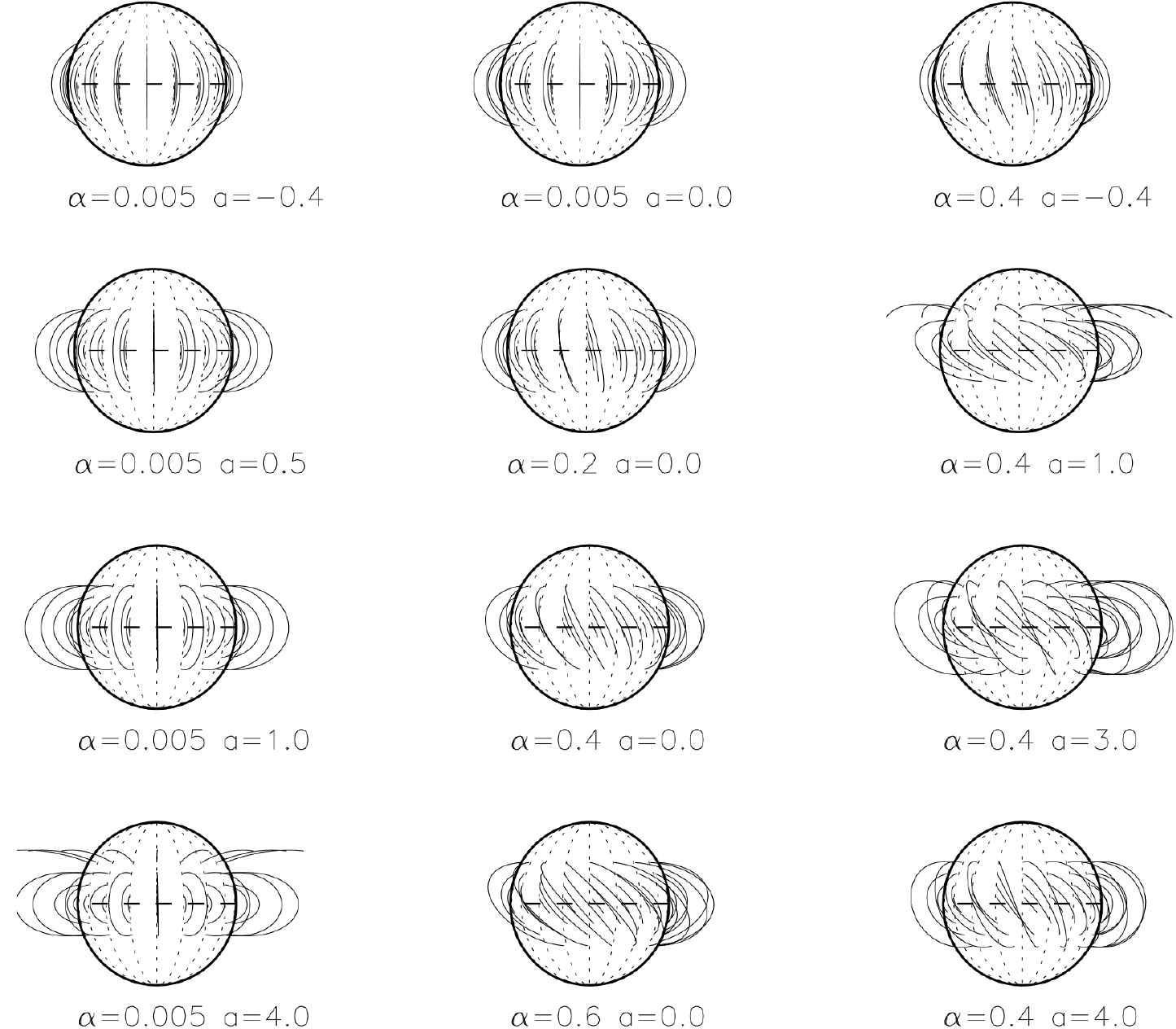}
\caption{Effect of the parameters $a$ and $\alpha$ in the MHS solution
  for a simple dipolar photospheric magnetic field. Closed magnetic
  field lines expand when $a$ is changed (left column), while the
  shear increases as the field-aligned current parameter $\alpha$ is
  increased (middle column). If both current systems act together
  (right column) then the field expands and becomes twisted. Image
  reproduced by permission from~\cite{2000ApJ...538..932Z}, copyright
  by AAS.}
\label{fig:zhao2000}
\end{figure}}

Another limitation of this magnetic field solution is that the magnetic energy is unbounded, since the Bessel functions decay too slowly as $r\rightarrow\infty$. But, as pointed out by \citet{2000ApJ...538..932Z}, the model is applicable only up to the cusp points of streamers, above which the solar wind outflow must be taken into account. So this problem is irrelevant in practical applications. For the $\alpha=0$ solution of \citet{1986ApJ...306..271B}, \citet{1994SoPh..151...91Z} showed how the model can be extended to larger radii by adding the ``current sheet'' extension of \citet{1971CosEl...2..232S} (described in Section~\ref{sec:pfss}). Instead of using an ``exterior'' solution in the external region they introduce an outer source surface boundary at $R_{\mathrm{ss}}\approx14\,R_{\odot}$, corresponding to the Alfv\'{e}n critical point. The resulting model better matches the shape of coronal structures and the observed IMF \citep{1995JGR...100...19Z} and is often termed the Current Sheet Source Surface (CSSS) model, though the authors term it HCCSSS: ``Horizontal Current Current Sheet Source Surface'', to explicitly distinguish it from the \citet{1971CosEl...2..232S} model using potential fields. This CSSS model has also been applied by \citet{2006AA...459..945S} to model the Sun's open magnetic flux (Section~\ref{sec:open}).

Finally, we note that \citet{2007AA...475..701W} have extended the numerical optimisation method (Section~\ref{sec:nlfff}) to magnetohydrostatic equilibria, demonstrating that it reproduces the analytical solution of \citet{1995AA...301..628N}. This numerical technique offers the possibility of more realistic pressure and density profiles compared to the analytical solutions, although there is the problem that boundary conditions on these quantities must be specified.

\subsection{Full magnetohydrodynamic models}
\label{sec:mhdfull}

In recent years, a significant advance has been made in the construction of realistic 
3D global  MHD models 
\citep{2006ApJ...653.1510R,2008ApJ...680..740D,2009ApJ...690..902L,2010ApJ...712.1219D,2012SoPh..279..207F}. Such models 
are required to give a self-consistent description of the interaction between the 
magnetic field and the plasma in the Sun's atmosphere.
They allow for comparison with observed plasma emission (Section~\ref{sec:mhd}), and enable more consistent modeling of the solar wind, so that the simulation domain may extend far out into the heliosphere \citep[e.g.,][]{2011SoPh..274..361R,2012JCoPh.231..870T,2012SoPh..279..207F}. On the other hand, additional boundary conditions are required (for example, on density or temperature), and the problem of non-uniqueness of solutions found in nonlinear force-free field models continues to apply here.
While these models are sometimes used to
simulate eruptive phenomena such as coronal mass ejections, in this review we consider only their application to non-eruptive phenomena. For applications to eruptive 
phenomena the reader is directed to the reviews of \cite{2006SSRv..123..251F} and \cite{2011LRSP....8....1C}.

\epubtkImage{mackay_fig9.png}{%
\begin{figure}[htbp]
\centering\includegraphics[scale=0.5]{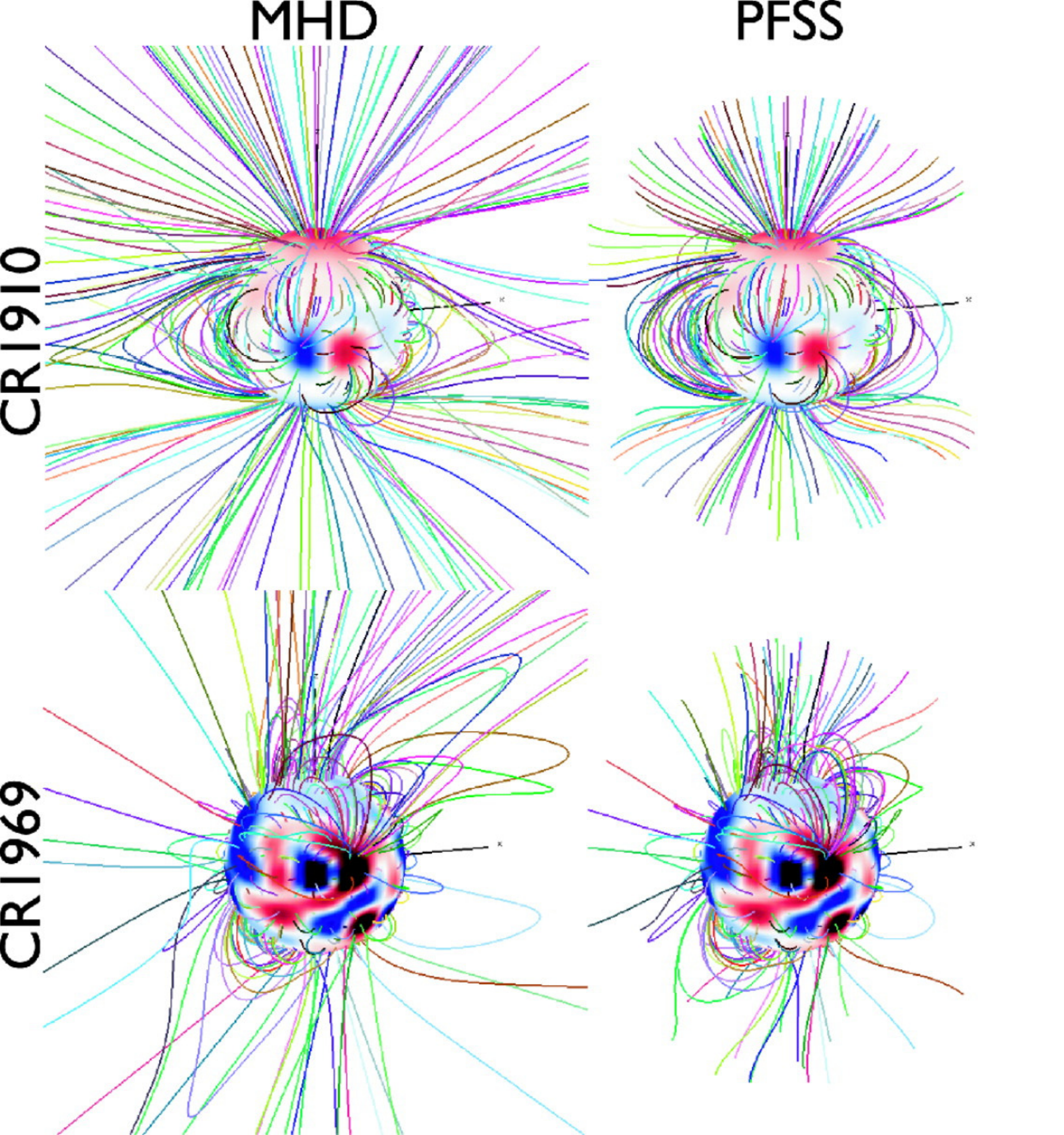}
\caption{Comparison of PFSS extrapolations (right column) with steady
  state MHD solutions (left column) for two Carrington
  rotations. Image reproduced by permission from Figure~5
  of~\cite{2006ApJ...653.1510R}, copyright by AAS.}
\label{fig:mackay9}
\end{figure}}

\newpage
An example of the resistive MHD equations used in the paper of \cite{2009ApJ...690..902L} for constructing a steady-state solution are:
\begin{eqnarray}
\frac{\partial \rho}{\partial t} + \nabla .\left( \rho \mathbf{v} \right) &=& 0, \label{eq:mhd1}\\
\rho \left(\frac{ \partial \mathbf{v}}{\partial t} + \mathbf{v}.\nabla\mathbf{v} \right) &=& 
  \frac{1}{c} \mathbf{J} \times \mathbf{B} - \nabla( p + p_w) + \rho \mathbf{g} + \nabla . (\nu \rho \nabla \mathbf{v}), \\
\frac{1}{\gamma -1} \left( \frac{\partial T}{\partial t} + \mathbf{v}. \nabla T \right) &=& - T \nabla .\mathbf{v} + \frac{m}{k_\rho} S, \\
S &=& - \nabla . \mathbf{q} - n_e n_p Q(t) + H_{\mathrm{ch}}, \label{eqn:ent}\\
 \nabla \times \mathbf{B} &=& \frac{ 4 \pi}{c} \mathbf{J}, \\ 
\nabla \times \mathbf{E} &=& -\frac{1}{c} \frac{ \partial \mathbf{B} }{\partial t}, \\
\mathbf{E} + \frac{\mathbf{v} \times \mathbf{B}}{c} &=& \eta \mathbf{J}, \label{eq:mhd2}
\end{eqnarray}
where $\mathbf{B}, \mathbf{J}, \mathbf{E}, \rho, \mathbf{v}, p, T,
\mathbf{g}, \eta, \nu, Q(t), n_e, n_p, \gamma=5/3, H_{\mathrm{ch}},
\mathbf{q}, p_w $ are the magnetic field, electric current density,
electric field, plasma density, velocity, pressure, temperature,
gravitational acceleration, resistivity, kinematic viscosity,
radiative losses, electron and proton number densities, polytropic index, coronal heating term, heat flux, and wave pressure, respectively. Different formulations of the MHD equations 
may be seen in the papers of \cite{2008ApJ...680..740D}, \cite{2010ApJ...712.1219D}, and \cite{2012SoPh..279..207F} 
where key differences
are the inclusion of resistive or viscous terms and the use of adiabatic or non-adiabatic energy equations.

To date, non-eruptive 3D global MHD simulations have been used to model the solar corona through two distinct
forms of simulation. Firstly, there is the construction of steady state coronal solutions from 
fixed photospheric boundary conditions 
\citep{2006ApJ...653.1510R,2008ApJ...682.1328V,2009ApJ...690..902L,2010ApJ...712.1219D}. 
To compute these solutions, the system is initialised by (i) specifying the photospheric distribution of flux 
(often from observations), (ii) constructing an initial potential magnetic field and, finally, (iii) superimposing 
a spherically symmetric solar wind solution. Equations~\eqref{eq:mhd1}\,--\,\eqref{eq:mhd2} (or their equivalents) are then integrated in time until a new equilibrium is 
found.  A key emphasis of this research is the direct comparison of the resulting 
coronal field and plasma emission with that seen in observations (Section~\ref{sec:mhd}).
In the paper of \cite{2008ApJ...682.1328V} a quantitative comparison of two global MHD models 
(Stanford: \citealp{2005ApJS..161..480H}, and Michigan: \citealp{2007ApJ...654L.163C}), with
coronal densities determined through rotational tomography was carried out. In general the models
reproduced a realistic density variation at low latitudes and below $3.5\,R_{\odot}$, however, both had 
problems reproducing the correct density in the polar regions.
In contrast to this construction of steady state MHD solutions, advances in computing power have recently enabled global non-eruptive MHD simulations with time-dependent photospheric boundary conditions.
These boundary conditions have been specified both in an idealised form \citep[e.g.,][]{2005ApJ...625..463L,2010ApJ...714..517E} and from synoptic magnetograms \citep{2011ApJ...731..110L}.
An initial application of these models has been to simulate the Sun's open flux and coronal holes: see Section~\ref{sec:ch}.

To quantify the difference between the magnetic field produced in global MHD simulations and PFSS extrapolations,
\cite{2006ApJ...653.1510R} carried out a direct comparison of the two techniques. The results from each technique are compared in Figure~\ref{fig:mackay9} for periods of low activity (top) and high activity (bottom).
 An important limitation of this study was that, for both the PFSS model and MHD model, only the
radial magnetic field distribution at the photosphere was specified in order to construct the coronal field.
Under this assumption, a good agreement was found between the two fields where the only differences were (i)
slightly more open flux and larger coronal holes in the MHD simulation, and (ii) longer field lines and more
realistic cusp like structures in the MHD simulations. While the two set of simulations closely agree, as discussed
by the authors it is important to note that if vector magnetic field measurements are applied on a global scale as an additional
lower boundary condition, then significant differences may result.

While global MHD models have mostly been applied to the Sun, \cite{2010ApJ...721...80C} applied the models to 
observations from AB~Doradus. Through specifying the lower boundary condition from a ZDI magnetic 
map and also taking into account the rotation (but not differential rotation) of the star the 
authors showed that the magnetic structure deduced from the MHD simulations was very different from that
deduced from PFSS models. Due to the rapid rotation a strong azimuthal field component
is created. They also considered a number of simulations where the base coronal
density was varied and showed that the resulting mass and angular momentum loss, which are
important for the spin-down of such stars, may be orders of magnitude higher than found for 
the Sun.

\newpage

\section[Application of Magnetic Flux Transport and Coronal Models]{Application\; of\; Magnetic\; Flux\; Transport\; and\; Coronal\; Models}
\label{sec:app}

In this section, we survey the combined application of magnetic flux
transport (Section~\ref{sec:mftm}) and coronal models (Section~\ref{sec:cf})
to describe a number of phenomena on the Sun. These 
include the Sun's open magnetic flux (Section~\ref{sec:open}), coronal holes
(Section~\ref{sec:ch}), and the hemispheric pattern of solar filaments
(Section~\ref{sec:fil}). Finally,
the application of global MHD models to predict the plasma emission from the Sun is
described in Section~\ref{sec:mhd}. It should be noted that when the magnetic
flux transport and coronal models are combined, two distinct modeling techniques
are applied, each of which have their own limitations. It is therefore important
to consider how the limitations of each affect the results. In addition to this, some models
also use observations as input. These observations also have many practical limitations
related to calibration, resolution, instrumental effects and data reduction effects. It is currently beyond 
the scope of the present article to consider the full consequences of these combined uncertainties
and limitations. However, all readers should keep this in mind when assessing results and outcomes.
In each of the following sections a brief description of the main properties of each
coronal phenomenon is first given, before discussing the application of the modeling techniques.

\subsection{Open flux}
\label{sec:open}

The Sun's open magnetic flux is the part of the Sun's large-scale magnetic field that
extends from the solar surface out into interplanetary space, where it forms what is known as the Interplanetary Magnetic Field (IMF). Understanding
the origin and variation of the open flux is important as:

\begin{itemize}
\item[-] It is the origin of the high speed solar wind and IMF, both of which directly 
interact with the Earth's magnetosphere.
\item[-] Irregularities or curvatures in the IMF modulate the flux of high energy Cosmic Rays
that impact the Earth's upper atmosphere. Such impacts produce \super{14}C 
\citep{1980Sci...207...11S} and \super{10}Be \citep{1990Natur.347..164B} isotopes which are found in
tree rings and ice cores, respectively. These isotopes may be used as historical data sets of solar 
activity since there is an anti-correlation between the strength of 
open flux and the abundance of the isotopes.
\item[-] The open flux has been correlated with cloud cover variations 
\citep{1997JASTP..59.1225S,1998PhRvL..81.5027S},  however, any relation to global warming is very 
uncertain.
\item[-] Strong correlations are found between the Sun's open flux and total solar irradiance \citep[TSI,][]{2002AA...382..678L}. Therefore historical data such as \super{14}C 
and \super{10}Be, which allow us to deduce the amount of open flux, 
may also be used as a proxy for past TSI variations.
\end{itemize}

\epubtkImage{mackay_fig4.png}{%
\begin{figure}[htbp]
\centering\includegraphics[scale=1]{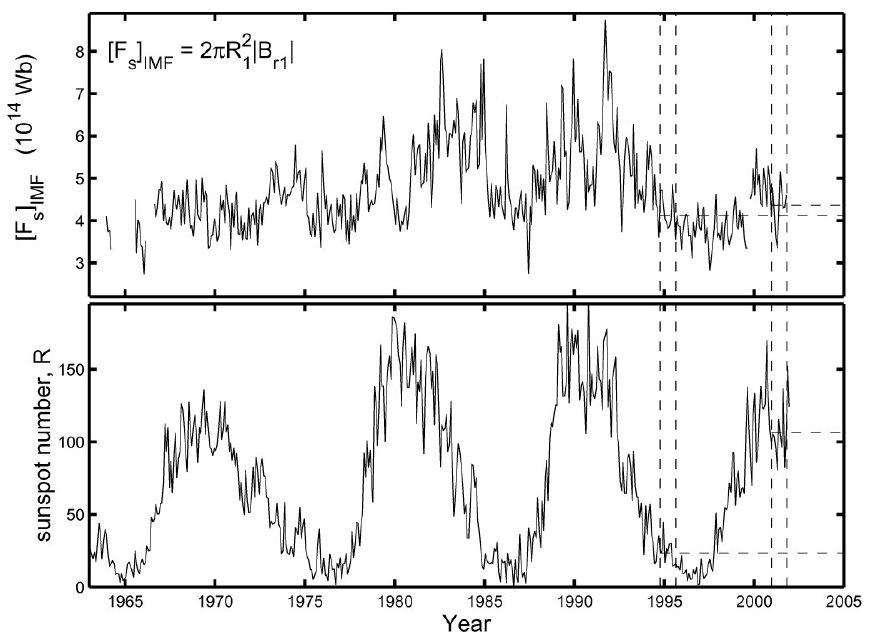}
\caption{Graph of open flux variation (top) and sunspot number
  (bottom) for cycles 20\,--\,23. Image reproduced by permission
  from~\cite{2004AnGeo..22.1395L}, copyright by EGU.}
\label{fig:obsopen}
\end{figure}}

Our present day understanding of the Sun's open flux comes mainly from an important result from
the Ulysses mission: the magnitude of the radial IMF component in the heliosphere 
is independent of latitude \citep{1995Sci...268.1007B}. Thus,
magnetometer measurements at a single location at 1~AU may be used to deduce the total open flux
simply by multiplying by the surface area of a sphere. 
In Figure~\ref{fig:obsopen}, the variation of open flux relative to sunspot number can be seen 
for the last 3.5 solar cycles. Over a solar cycle the open flux varies at 
most by a factor of two, even though the surface flux shows a much larger
variation. However, this 
variation is not regular from one cycle to the next. It is clear from the graph 
that Cycles~21 and 22 have a much larger variation in open flux than Cycles~20 and 23, 
indicating a complex relationship between the surface and open flux. A key property of 
the open flux is that it slightly lags behind the variation in sunspot number and peaks 1\,--\,2~years after cycle maximum. The peak time of open flux therefore occurs around the same time as polar field reversal, indicating that contributions from both polar coronal holes and low-latitude coronal holes must be important.

While both direct and indirect observations exist for the Sun's open flux, such measurements 
cannot be made for other stars. Due to this, theoretical models which predict the magnitude 
and spatial distribution of open
flux are used. Within the stellar context, the distribution of open flux with latitude 
\citep{2006MNRAS.367..592M} plays a key role in determining the mass and angular
momentum loss \citep{2007ApJ...654L.163C} and subsequently the spin down of stars 
\citep{1967ApJ...148..217W,2003ApJ...590..493S,2007AA...463...11H}.

Over the last 20 years a variety of techniques has been developed to model the
origin and variation of the Sun's open flux. 
One category are the semi-empirical magnitude variation models. These follow only the total open flux, and are driven by observational data. Different models use either geomagnetic data from Earth \citep{1999Natur.399..437L,2009ApJ...700..937L}, sunspot numbers \citep{2000Natur.408..445S,2002AA...383..706S}, or coronal mass ejection rates \citep{2011JGRA..11604111O}. Such models have been used to extrapolate the open flux back as far as 1610 (in the case of sunspot numbers), although the results remain uncertain. However, these models consider only the total integrated open flux, not its spatial distribution and origin on the Sun. To describe the latter, coupled photospheric and coronal magnetic field models have been applied.
The photospheric field is specified either from synoptic magnetic field observations or from magnetic flux transport simulations (Section~\ref{sec:mftm}), and a wide variety of coronal models have been used. These range from potential 
field source surface (PFSS, Section~\ref{sec:pfss}) models to current sheet source 
surface (CSSS, Section~\ref{sec:pfss}) models and more 
recently to nonlinear force-free (NLFF, Section~\ref{sec:nlfff}) models. In all of 
these coronal models, field lines
reaching the upper boundary are deemed to be open, and the total (unsigned) magnetic flux through this boundary represents the open flux.

\enlargethispage{\baselineskip}
\citet{2002JGRA..107.1302W} combined synoptic magnetograms and PFSS 
models to compute the open flux from 1971 to 1998. The synoptic magnetograms
originated from either Wilcox Solar Observatory (WSO, 
1976\,--\,1995) or Mount Wilson Observatory (MWO, 1971\,--\,1976, 1995\,--\,1998). The authors found that
to reproduce a good agreement to IMF field measurements at 1~AU, they had to multiply the
magnetograms by a strong latitude-dependent correction factor ($4.5\mbox{\,--\,}2.5 \sin^2 \lambda$). 
This factor, used to correct for saturation effects in the magnetograph observations, 
significantly enhanced the low latitude fields compared to the high latitude fields but
gave the desired result. 

\epubtkImage{mackay_fig5.png}{%
\begin{figure}[htbp]
\centering\includegraphics[width=\textwidth]{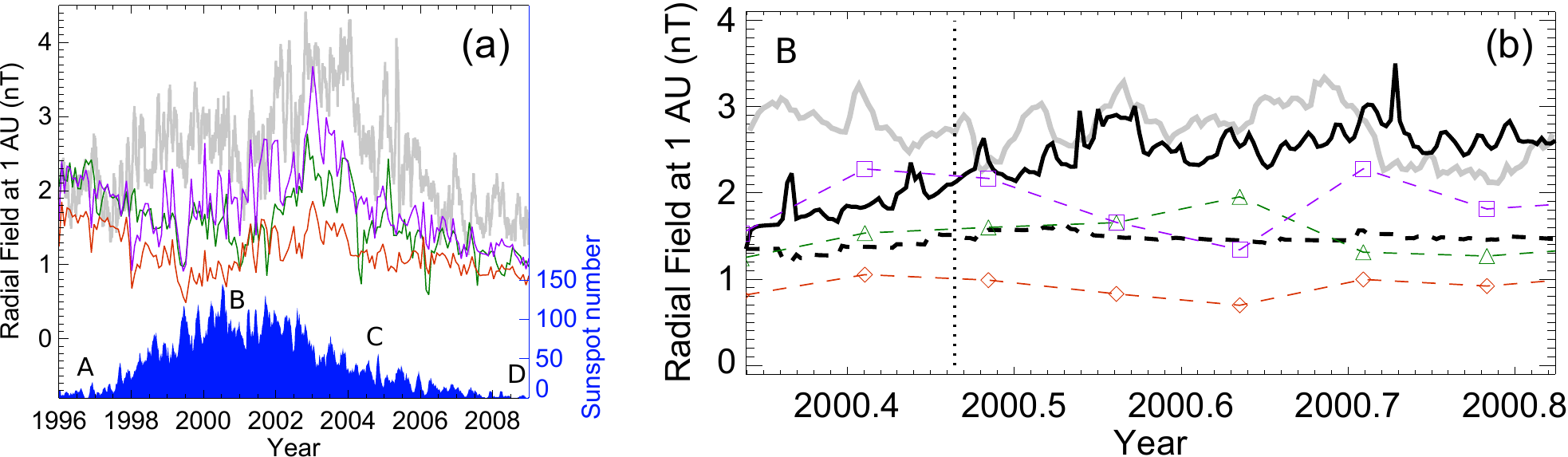}
\caption{Graph of (a)~IMF variation (grey line) and PFSS
  approximations to the IMF field (collared lines). (b)~Non-potential
  open flux estimate (thick black line) along with IMF variation (grey
  line) and PFSS approximations to the IMF (coloured lines) over a
  6~month period. Image reproduced by permission
  from~\cite{2010JGRA..11509112Y}, copyright by AGU.}
\label{fig:theopen}
\end{figure}}

The use of such a latitude dependent correction factor was recently questioned by 
\cite{2007ApJ...667L..97R}, who pointed out that the correction factor used by 
\cite{2002JGRA..107.1302W} was only correct for use on MWO data and did not 
apply to WSO data. Instead WSO data, which made up the majority of 
the magnetogram time series used, should have a constant 
correction factor of 1.85 \citep{Svalgaard2006}. On applying this appropriate correction
and repeating the calculation, 
\cite{2007ApJ...667L..97R} showed that PFSS models gave a poor fit to observed IMF 
data \citep[see Figure~4 of][]{2007ApJ...667L..97R}. To resolve the difference, 
\cite{2007ApJ...667L..97R} put forward an alternative explanation. He assumed that the open flux
has two contributions: (i) a variable background contribution based on the
photospheric field distribution on the Sun at any time, such as that obtained from the PFSS 
model, and (ii) a short term enhancement due to interplanetary coronal mass ejections (ICMEs),
for which he gave an order-of-magnitude estimate. Through combining the two a good agreement to IMF field observations was found. 

As an alternative to synoptic magnetic field observations, magnetic flux transport models 
have been widely used to provide the lower boundary condition in studies of the origin and 
variation of the Sun's open flux. A key element in using magnetic flux transport
simulations is that the distribution and strength of the high latitude field are
determined by the properties and subsequent evolution of the bipoles which emerge at lower 
latitudes. Hence, assuming the correct input and advection parameters, the model will
produce a better estimate of the polar field strength compared to that found in
magnetogram observations which suffer from severe line-of-sight effects above 60\textdegree\
latitude. Initial studies (using a PFSS coronal model) 
considered the variation in open flux from a single bipole 
\citep{2000GeoRL..27..621W,2002SoPh..207..291M} as it was advected across the solar surface. 
Through doing so, the combined effects of latitude of emergence, Joy's Law and
differential rotation 
on the open flux was quantified. Later studies extended these simulations to include the full solar 
cycle, but obtained conflicting results. 

To start with, \cite{2002SoPh..209..287M}, using idealised simulations and commonly used
transport parameters,  found that such 
parameters when combined with PFSS models produced an incorrect time-variation of open flux. The open 
flux peaked at cycle minimum, completely out of phase with the solar cycle and inconsistent with 
the observed 1\,--\,2~year lag behind solar maximum. The authors attributed this to the fact that in PFSS models only 
the lowest order harmonics (see Section~\ref{sec:pfss}), which are strongest at cycle minimum and weakest at maximum, 
significantly contribute to the open flux. They concluded that the only way to change this was 
to include non-potential effects (see below) which would increase the open flux contribution from 
higher order harmonics during cycle maximum.  In response, \cite{2002ApJ...580.1188W} repeated the 
simulation and found similar results when using bipole fluxes and parameters deduced from observations.
However, they were able to obtain the correct variation in open flux by both trebling all 
observed bipole fluxes and increasing the rate of meridional flow to 25~\ms
(dash-dot-dot-dot line in Figure~\ref{fig:ftpro}b).
These changes had the effect of significantly enhancing the low latitude field at cycle maximum 
but weakening the polar fields at cycle minimum. 
As a result the correct open flux variation was found. 
Opposing such a strong variation in the input parameters, \cite{2006AA...459..945S}
put forward another possibility. Through modifying the magnetic flux transport model to include a radial decay term for $B_r$ (Section~\ref{sec:mftm}), and changing the coronal model to 
a CSSS model with $R_{\mathrm{cp}} \sim 1.7\,R_{\odot}$ and $ R_{\mathrm{ss}} \sim 10\,R_{\odot}$ (Section~\ref{sec:mhs}), they 
found that the correct 
open flux variation could be obtained, but only if new bipole tilt angles were decreased from 
$0.5 \lambda$ to $0.15 \lambda$. Later studies by \cite{2010ApJ...719..264C} using the same technique 
showed that the radial decay term was not required if the tilt angles of the bipoles varied 
from one cycle to the next, with stronger cycles having weaker tilt angles.
 
The discussion above shows that, similar to the polar field strength (Section~\ref{sec:mftm}), the correct variation of the Sun's open flux may be obtained through a
variety of methods. These include variations in the bipole tilt angles, rates of meridional flow,
and different coronal models. More recently, \cite{2010JGRA..11509112Y} showed that standard values for
bipole tilt angles and meridional flow may produce the correct variation of open flux if a 
more realistic physics-based coronal model is applied (see Section~\ref{sec:ftmf}). By allowing for 
electric currents in the corona, a better
agreement to the IMF measurements can be found compared to those from PFSS models.
Figure~\ref{fig:theopen}a compares various PFSS extrapolations using different magnetogram data 
(coloured lines) to the measured IMF field (grey line). In this graph the discrepancy between all 
potential field models and the IMF is particularly apparent around cycle maximum. 
To resolve 
this, \cite{2010JGRA..11509112Y} ran the global non-potential model described in Section~\ref{sec:ftmf}
over four distinct 6-month periods (labeled A\,--\,D in Figure~\ref{fig:theopen}a).
Two periods were during low solar activity (A and D) and two during high solar activity (B and C). 
The results of the 
simulation for period B can be seen in Figure~\ref{fig:theopen}b. In the plot, 
the dashed lines denote open flux from various PFSS extrapolations, the grey line the 
observed IMF field strength and the black solid line the open flux from the non-potential global 
simulation. This clearly gives a much better agreement in absolute magnitude terms
compared to the PFSS models.  
\cite{2010JGRA..11509112Y} deduced that the open flux in their model has three main sources. The first is a 
background level due to the location of the flux sources. This is the component that may be captured
by PFSS models, or by steady-state MHD models initialised with PFSS extrapolations \citep{2012SoPh..279..207F}. The second is an enhancement due to
electric currents, which results in an inflation 
of the magnetic field, visible in Figure~\ref{fig:theopen}b as an initial steady increase of the open flux curve during the first month to a higher base level. This inflation is the result of large scale flows such as
differential rotation and meridional flows along with flux emergence, reconfiguring the coronal field. A similar inflation due to electric currents enhances the open flux in the CSSS model, although there the pattern of currents is arbitrarily imposed, rather than arising physically.
Finally, there is a sporadic component to the open flux as a result of flux rope 
 ejections representing CMEs similar to that proposed by \cite{2007ApJ...667L..97R}.  While this model gives 
 one explanation for the open flux shortfall in PFSS models, we note that alternative explanations have been put 
forward by a variety of authors \citep{2009ApJ...700..937L,2006JGRA..11109115F}. Again, although models have been successful in explaining the variation of the open flux, there is still an uncertainty on the true physics required. Free parameters that presently allow for multiple models must be constrained by future observations. At the same time, it is evident from using existing observations that care must be taken in their use. In particular, the effect on the modeling results of practical limitations such as calibration, resolution, instrumental effects, and data reduction effects must be quantified.

Although they include the spatial distribution of open flux, the models described above have mainly focused on the total amount of open flux, because this may be compared with in situ measurements. However, these models can also be used to consider how open flux evolves across the solar surface, and its relationship to coronal holes.
Presently, there are two opposing views on the nature of the evolution of open flux.
The first represents that of \cite{2001ApJ...560..425F} who suggest that open flux may diffuse across the solar surface through interchange reconnection. In this model, the open flux may propagate into and  
through closed field regions.  The second view represents that of
\cite{2007ApJ...671..936A} who postulate that open flux can never be isolated and thus 
cannot propagate through closed field regions. Results of global MHD simulations testing these ideas are 
discussed in the next section.

\epubtkImage{wang_ch.png}{%
\begin{figure}[htbp]
\centering\includegraphics[scale=0.8]{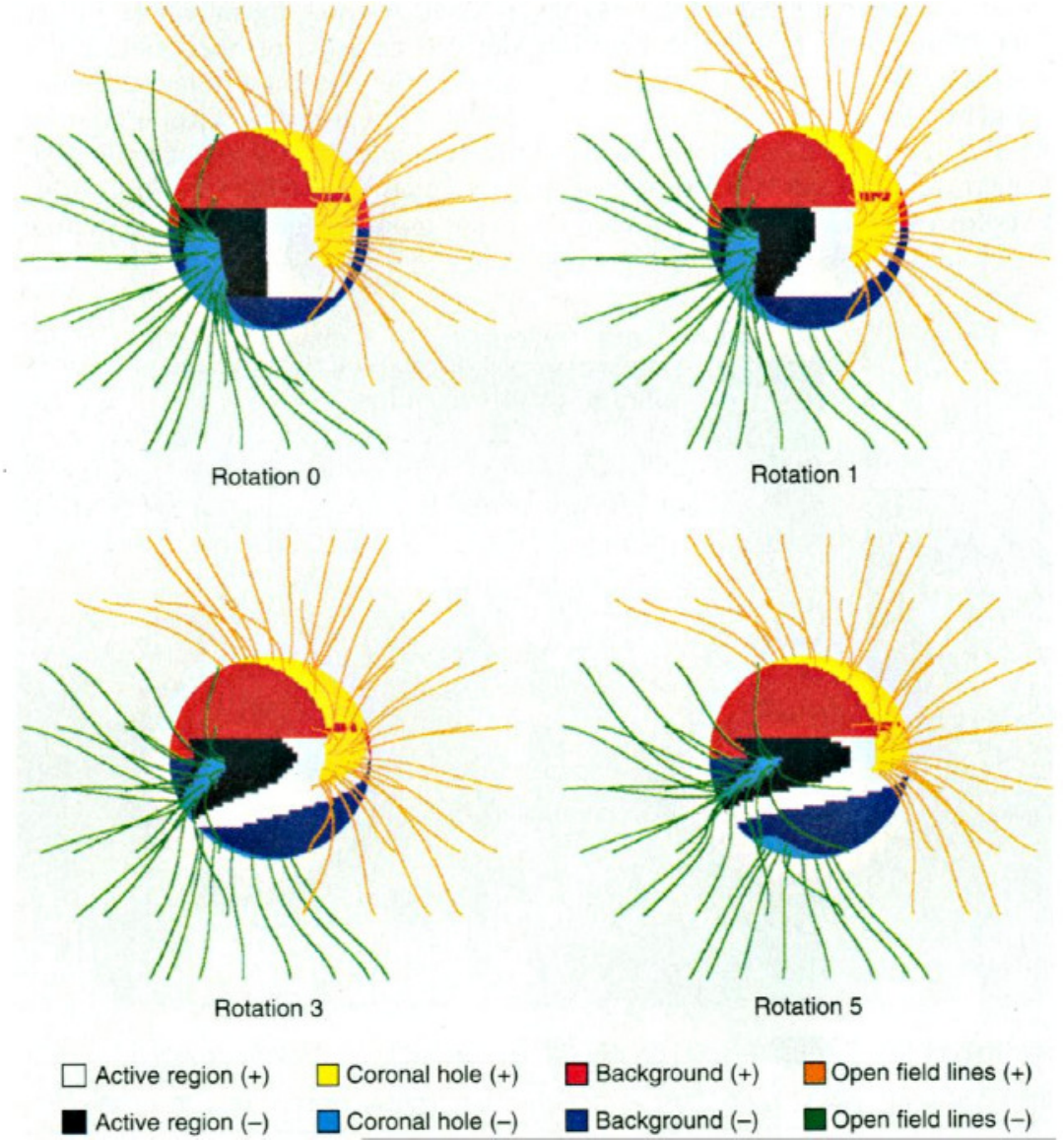}
\caption{Effect of differential rotation on open field regions for a
  bipole (black/white) located in a dipolar background field (red/dark
  blue). The open region (``coronal hole'') in the North (yellow)
  rotates almost rigidly, while that in the South (light blue) is
  sheared. Image reproduced by permission
  from~\cite{1996Sci...271..464W}, copyright by AAAS.}
\label{fig:corhole}
\end{figure}}

\subsection{Coronal holes}
\label{sec:ch}

Coronal holes are low density regions of the corona which appear dark in X-rays or EUV, or as light, blurred patches in He\,{\sc i}~10830~\AA\ \citep{2009LRSP....6....3C}. It has been known since the \emph{Skylab} era of the early 1970s that coronal holes are associated with regions of open magnetic field lines (Section~\ref{sec:open}), along which plasma escapes into the heliosphere \citep{1977RvGSP..15..257Z}. They therefore play an important role in understanding the sources of different solar wind streams and the relationship between the Sun and the heliosphere.

The basic link between coronal holes and open magnetic field was found using PFSS extrapolations from synoptic magnetogram observations. While the open field regions in such extrapolations produce pleasing agreement with observed coronal hole boundaries \citep{1982SoPh...79..203L,1996Sci...271..464W,1998JGR...10314587N,2002JGRA..107.1154L}, our understanding has been enhanced by coupling coronal magnetic models with photospheric simulations, in the following ways:
\begin{enumerate}
\item Photospheric simulations can be used to produce a more accurate picture of the global magnetic field by assimilating observed data and following the evolution on timescales shorter than the 27~days of the synoptic charts. \citet{2003SoPh..212..165S} found that some open field regions in such a model were quite stable for weeks or even months, whereas others fluctuated rapidly.
\item The photospheric evolution can be imposed or controlled, in order to test hypotheses for the behaviour of coronal holes, and to try to understand the origin of their observed properties.
\end{enumerate}

We outline here the studies that fall into the second category. Firstly, a series of papers by Wang, Sheeley, and co-workers, coupling surface flux transport to PFSS extrapolations, has significantly added to our understanding of coronal hole dynamics. By isolating different terms in the flux transport model, \citet{1990ApJ...365..372W} demonstrate how each of the terms in the flux transport model play their roles in producing the global pattern of coronal holes. Supergranular diffusion spreads out active region flux so as to create unipolar areas containing open flux, and also facilitates the build up of trailing polarity holes in each hemisphere after Solar Maximum by annihilating leading polarity flux at the equator. Differential rotation helps to accelerate the formation of axisymmetric polar holes by symmetrising the active region flux distribution. Meridional flow (i) concentrates flux at the poles -- preventing the polar holes from spreading to lower latitudes, (ii) hastens the decay of low-latitude holes by transporting them to mid-latitudes where differential rotation is more efficient, and (iii) impedes cancellation across the equator, thus reducing the flux imbalance in each hemisphere. 

A striking observation arising from \emph{Skylab} was that coronal holes rotate differently from the underlying photosphere. This is exemplified by the \emph{Skylab} ``Coronal Hole 1''. \citet{1988SoPh..117..359N} were able to reproduce and explain the behaviour of this hole using the coupled flux transport and PFSS model \citep[see also][]{1993ApJ...414..916W}. The mechanism is illustrated by their model of a single bipolar active region in a dipolar background field \citep[][reproduced here in Figure~\ref{fig:corhole}]{1990ApJ...365..372W,1996Sci...271..464W}. There are two important points. The first is that coronal hole boundaries in the PFSS model are determined by the \emph{global} magnetic structure of the coronal field. This is because the source surface field depends only on the lowest order spherical harmonics. In particular, the open field regions in Figure~\ref{fig:corhole} are not simply a superposition of those that would be obtained from the bipole and background fields individually \citep{1996Sci...271..464W}. Hence, it need not be surprising that coronal holes are observed to rotate differently from the footpoints of individual field lines. The second point is that the rotation rate is determined by the non-axisymmetric component only. In other words, the ``coronal hole extension'' in Figure~\ref{fig:corhole} must rotate approximately with the bipole giving rise to it. It follows that the rotation rate of the coronal hole matches that of the photospheric latitude where the bipole is located, not necessarily the latitude where the open field lines have their footpoints. This explains why the northern and southern coronal hole extensions in Figure~\ref{fig:corhole} behave differently: the bipole flux causing the northern hole extension is located in a narrow band close to the equator, so that the hole rotates rigidly with the equatorial 27-day period. On the other hand, the flux causing the southern hole extension is spread between 0\textdegree\ and 40\textdegree\ latitude, so that this hole is considerably sheared by differential rotation. The same idea explains why rigidly rotating coronal holes are more prevalent in the declining phase of the solar cycle: during this phase, the non-axisymmetric flux is concentrated more at lower latitudes.

More recently, new insights into coronal hole evolution have started to come from coupling global MHD models (Section~\ref{sec:mhdfull}) to surface flux transport. For example, an important implication of the PFSS model for coronal magnetic fields is that continual magnetic reconnection is necessary to maintain the current-free field \citep{2004ApJ...612.1196W}. This has been tested in a global MHD model by \citet{2005ApJ...625..463L}, who applied differential rotation to a configuration consisting of a single bipole in a dipolar background field. They confirm the results of \citet{1996Sci...271..464W} that the coronal hole extension rotates nearly rigidly without significant change, even when the surface field is significantly sheared by differential rotation. The dominant reconnection process was found to be interchange reconnection, with continual reconnection opening field on the eastern hole boundary and closing it on the western boundary. In some cases, when field line footpoints pass over the coronal hole boundaries multiple times, multiple openings and closing occur. However, it should be noted that these simulations used a uniform resistivity orders of magnitude greater than that in the real corona, so that reconnection may be less efficient in reality. \citet{1996JGR...10115547F} has proposed an alternative scenario where open field lines continually circulate in latitude and longitude as their footpoints rotate differentially. This process was applied to explain the origin of high latitude Solar Energetic Particles (SEPs) seen by Ulysses \citep{1995Sci...268.1013K,1995GeoRL..22.3361M}, which are thought to originate in low latitude corotating interacting regions \citep[CIRs,][]{1976ApJ...203L.149M}. However, in a test with their global MHD model and an initial tilted dipole field, \citet{2006ApJ...642L..69L} found that although field lines did move in latitude, the coronal hole boundaries rotated in a manner consistent with the extrapolation models, and not perfectly rigidly as predicted by \citet{1996JGR...10115547F}. This meant that the latitudinal excursion of around 25\textdegree\ found in the simulations was insufficient to explain that required by \citet{1996JGR...10115547F}. 

\epubtkImage{dhm_openflux.png}{%
\begin{figure}[htb]
  \centering\includegraphics[scale=0.6]{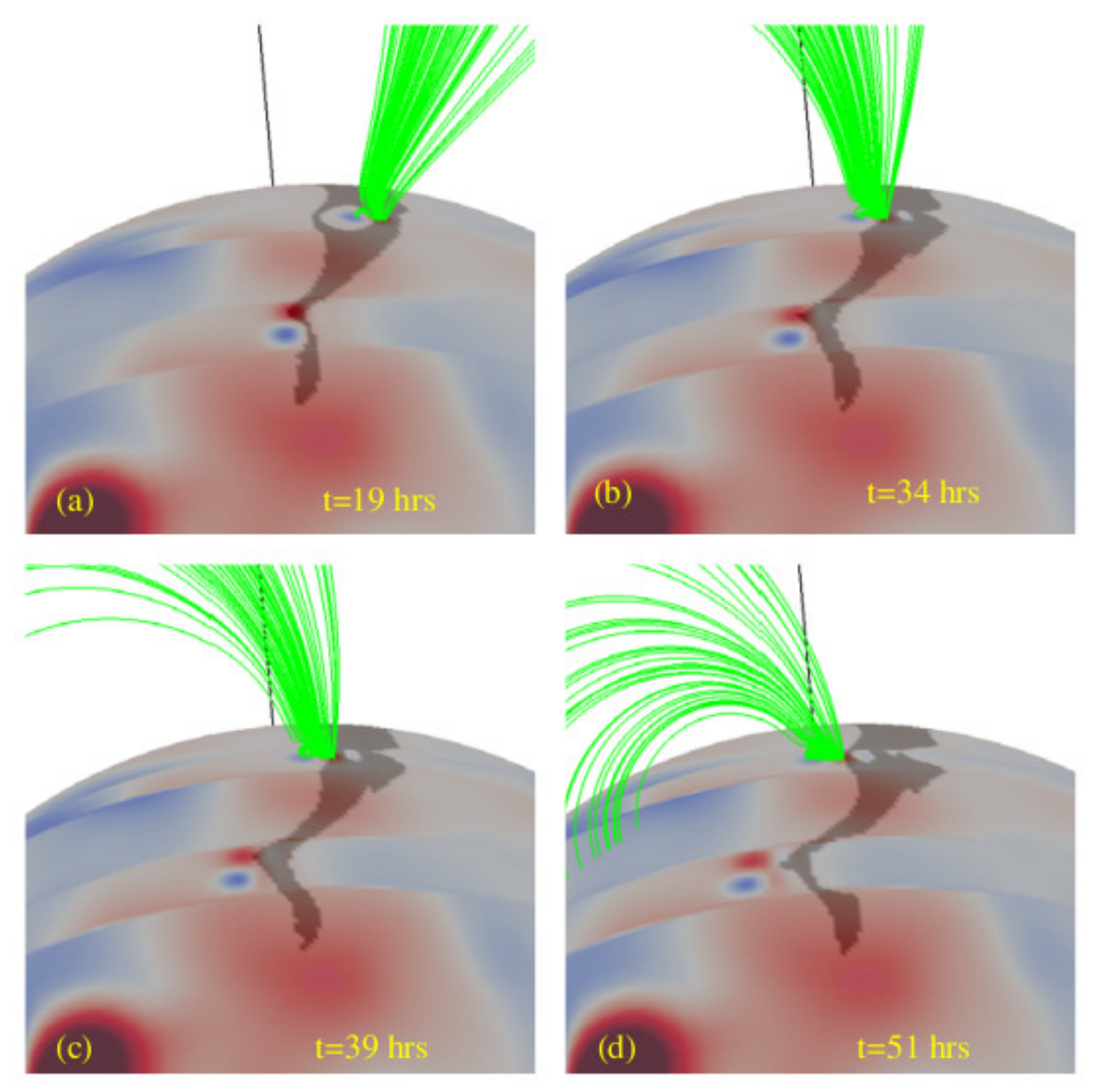}
\caption{Simulation results from the paper of
  \citet{2011ApJ...731..110L} where a bipole is advected from an open
  field region (shaded area on the solar surface) to a closed field
  region across the boundary of a coronal hole. Red denotes positive
  photospheric flux and blue negative flux. The green lines denote (a)
  initially open field lines of the bipole which then successively
  close down (b)\,--\,(d) as the bipole is advected across the coronal
  hole boundary. Image reproduced by permission, copyright by AAS.}
\label{fig:final}
\end{figure}}

Recently, \citet{2011ApJ...731..110L} have applied their global MHD model to test another hypothesis, namely, that the total amount of open flux is conserved. This has been proposed by Fisk and co-workers based on heliospheric observations \citep{2001ApJ...560..425F}. They propose that open flux is transported by interchange reconnection (between open and closed field lines), but not created or destroyed through other forms of reconnection.  A feature of this model is that open field lines may be transported through closed field regions in a random manner during the reversal of the polar field. In their simulations, \citet{2011ApJ...731..110L} start with a synoptic magnetogram for CR1913, but add two small bipoles of varying orientation inside one of the large equatorward coronal hole extensions.  They then advect the bipoles through the coronal hole from the open to closed field regions. The idea is to test (a) whether open flux associated with the bipoles can be transported out of the coronal hole and into the closed field region \citep[as required by the interchange model;][]{2006JGRA..11109115F}, and (b) whether isolated coronal holes can exist in each hemisphere. The 
results (Figure~\ref{fig:final}) seem to contradict the requirements of the interchange model, and to uphold the conjecture of 
\citet{2007ApJ...671..936A}, namely that what appear to be isolated coronal holes in each hemisphere, are in fact, always 
connected (by very thin corridors) of open flux to the polar holes in each hemisphere. These thin corridors may 
intersect with strong field regions that form part of the streamer belt. Thus, they may have a non-negligible contribution
to the amount of open flux, which may help resolve the discrepancy between the amount of open magnetic flux obtained in models
and that directly observed through IMF measurements (Section~\ref{sec:open}).
Following this, \cite{2011ApJ...731..111T} constructed a topological model for the open flux corridors that connect coronal holes. Further studies by \cite{2009ApJ...707.1427E,2010ApJ...714..517E} considered idealised case studies of 
how magnetic reconnection affects the boundaries of coronal holes when either a bipole is (i) 
advected across the 
coronal boundary \citep{2010ApJ...714..517E} or (ii) stressed by rotational motions \citep{2009ApJ...707.1427E}.
In line with the conjecture of \citet{2007ApJ...671..936A} and quasi-steady model results of 
\citet{2011ApJ...731..110L}, the authors see a smooth transition of the open flux where the open field regions are 
never isolated. Such evolution of coronal hole boundaries along with the translation and opening/closing 
of magnetic field lines has important consequences for the solar wind. A full discussion of this is beyond the 
scope of the present review but details can be found in \cite{2009LRSP....6....3C} and \cite{2011SSRv..tmp..148A}.

\subsection{Hemispheric pattern of solar filaments} 
\label{sec:fil}

Solar filaments (a.k.a.\ prominences) are cool dense regions of plasma ($T \sim 10,000\mathrm{\ K}$) 
suspended in the hotter ($T \sim 10^{6}\mathrm{\ K}$) solar corona. Coronal magnetic fields 
are key to the existence of solar filaments, where they provide both support against gravity and insulation 
of the cool material from the surrounding hot corona. The main observational properties  of filaments and 
the theoretical models used to describe them are described  in the reviews by \cite{2010SSRv..151..243L} 
and \cite{2010SSRv..151..333M}. 

In recent years, solar filaments and their birth grounds called filament channels 
\citep{1998ASPC..150..257G}, which always overlie photospheric polarity inversion lines (PILs),
have been classified by the orientation of their main axial magnetic field. 
This orientation, named the filament chirality \citep{1994ssm..work..303M} may take one of two 
forms: dextral or sinistral. Dextral/sinistral filaments have an axial magnetic field that 
points to the right/left when the main axis of the filament is viewed from the positive polarity 
side of the PIL.  In force-free field models 
\citep{1998AA...329.1125A,1999SoPh..185...87M,2000ApJ...539..983V,2005ApJ...621L..77M} this chirality 
is directly related to the dominant sign of magnetic helicity that is contained within the filament and
filament channel. A dextral filament will contain dominantly negative helicity, 
a sinistral filament positive helicity. As filaments and their channels 
form over a wide range of latitudes on the Sun, ranging from the active belts to the polar crown, they
may be regarded as useful indicators of sheared non-potential fields  and magnetic helicity within the 
solar corona.

A surprising feature of the chirality of filaments is that it displays a large-scale hemispheric pattern: 
dextral/sinistral filaments dominate in the 
northern/southern hemispheres, respectively \citep{1994ssm..work..303M,1997SoPh..175...27Z,2003ApJ...595..500P,2007SoPh..245...87Y}. This
pattern is intriguing as it is exactly opposite to that expected from
differential rotation acting on a North-South coronal arcade. Although
dextral/sinistral filaments dominate in the northern/southern
hemisphere, observations show that exceptions to this pattern do
occur. Therefore any model which tries to explain the formation of
filaments and their dominant axial magnetic fields must explain not
only the origin of this hemispheric pattern but also why
exceptions to it arise.

Since filament chirality is directly related to magnetic helicity 
(dextral $\sim$ negative, sinistral $\sim$ positive) the formation and transport of 
filaments across the solar surface is an indication of the global pattern of magnetic 
helicity transport  on the Sun 
\citep{2008ApJ...680L.165Y}, a key feature in explaining many eruptive phenomena. 
Static extrapolation techniques (including MHD models of the relaxation type) cannot study the generation nor transport of this helicity 
across the surface of the Sun. This can only be achieved with models that couple the evolution of 
both photospheric and coronal fields over long periods of time. 

The first attempt to explain the hemispheric pattern of filaments using global magnetic
flux transport models was carried out by \cite{1998ApJ...501..866V}.  By using observed 
magnetic flux distributions and initial coronal fields which were potential, they simulated the 
non-equilibrium evolution of the coronal field. They found that
the flux transport effects acting alone on potential fields
create approximately equal numbers of dextral and sinistral channels
in each hemisphere, in contradiction with observations.  Following this, \cite{2005ApJ...621L..77M}
re-considered the origin of the hemispheric pattern through combined
flux transport and magneto-frictional relaxation simulations (Section~\ref{sec:ftmf}), 
where the coronal field responds to surface motions by relaxing to a
nonlinear force-free equilibrium. In the simulations the
authors considered an idealised setup of initially isolated
bipoles. Therefore, in contrast to the study of \cite{1998ApJ...501..866V}, 
they did not use initial potential fields. The authors demonstrate that the
hemispheric pattern of filaments may be explained through the
observational properties of newly-emerging bipoles such as (i) their
dominant tilt angles \citep[--10\textdegree:30\textdegree,][]{1989SoPh..124...81W}, and (ii) the
dominant helicity with which they emerge in each hemisphere
\citep{1995ApJ...440L.109P}. A key feature of these simulations was that the possible occurrence of exceptions to the hemispheric pattern was quantified for the first time, arising from large positive bipole tilt
angles and minority helicity.

In a more conclusive study, the results of \cite{2005ApJ...621L..77M} have been tested 
through a direct comparison between theory and observations by
\cite{2007SoPh..245...87Y,2008SoPh..247..103Y}. First, they used H$\alpha$ observations
from BBSO over a 6 month period to determine the location and
chirality of 109 filaments \citep{2007SoPh..245...87Y} relative to the
underlying magnetic flux.  In the second stage they used combined
magnetic flux transport and magneto-frictional simulations (Section~\ref{sec:ftmf}), 
based on actual photospheric magnetic
distributions found on the Sun. Unlike previous
studies, these were run for the whole six month period without
ever resetting the surface field back to that found in observations or
the coronal field to potential. To maintain accuracy over this time, 119 bipoles had to be emerged 
with properties determined from observations. Hence, the simulations were able to
consider long term helicity transport across the solar surface from
low to high latitudes.
\cite{2008SoPh..247..103Y} carried out a direct
one-to-one comparison of the chirality produced by the model with the
observed chirality of the filaments. An example of this can be seen in
Figure~\ref{fig:mackay_extra2}, where Figure~\ref{fig:mackay_extra2}a
shows the global field distribution after 108~days of
evolution. The enlargement in Figure~\ref{fig:mackay_extra2}b shows a simulated flux rope 
structure with an axial field of dextral
chirality. The real H$\alpha$ filament that formed at this
location can be seen in Figure~\ref{fig:mackay_extra2}c.
Through studying the barbs and applying a
statistical technique, the filament was determined to be of dextral
chirality so the chirality formed in the simulation matches that of
the filament.

Through varying the sign and amount of helicity emerging within the
bipoles, \cite{2008SoPh..247..103Y} (see their Figure~5b) show that by
emerging dominantly negative helicity in the northern hemisphere and
positive in the southern, a 96\% agreement can be found between the
observations and simulations. An important feature is that the agreement is equally good for
minority chirality filaments as well as for dominant chirality
filaments.  Therefore, the global coronal model applied was equally good in producing positive and negative helicity
regions at the correct times and spatial locations to represent solar filaments.
Another  key feature of the simulations is that a better
agreement between the observations and simulations is found the longer
the simulations are run.  This indicates that the Sun has a long term
memory of the transport of helicity from low to high latitudes. The
reason for this high agreement is described in the paper of
\cite{2009SoPh..254...77Y} where seven different mechanisms are involved in
producing the observed chiralities. The results demonstrate that the
combined effects of differential rotation, meridional flow,
supergranular diffusion, and emerging twisted active regions are
sufficient to produce the observed hemispheric pattern of filaments.
While the model of \cite{2008SoPh..247..103Y} obtained an excellent agreement
with the observed  chirality of filaments, the 6 month simulation was unable to
reproduce dextral/sinistral chirality along the polar crown PILs in the northern/southern
hemispheres. The cause was that the simulation was not run for long enough to allow meridional
flow, which acts on a time scale of 2~yr (see Section~\ref{sec:basicmftm}), to transport the 
helicity produced at low latitudes up into the polar regions.  A later simulation running for a full 11-year 
solar cycle \citep[see Figure~4a of][]{2012ApJ...753L..34Y} 
was able to obtain the correct dextral/sinistral chirality  along the polar crown PILs in each hemisphere, once
the duration of the simulation exceeded the meridional flow timescale.  
This shows that, at least in the context of solar filaments, the flux transport and magneto-frictional model adequately describes the build-up and transport of helicity from low to high latitudes on the Sun. The model has subsequently been applied to the formation of magnetic flux ropes and their subsequent loss of equilibrium \citep{2009ApJ...699.1024Y,2010ApJ...709.1238Y}, a possible mechanism for the initiation of coronal mass ejections.

\epubtkImage{mackay_fig6.png}{%
\begin{figure}[htbp]
\centering\includegraphics[scale=0.5]{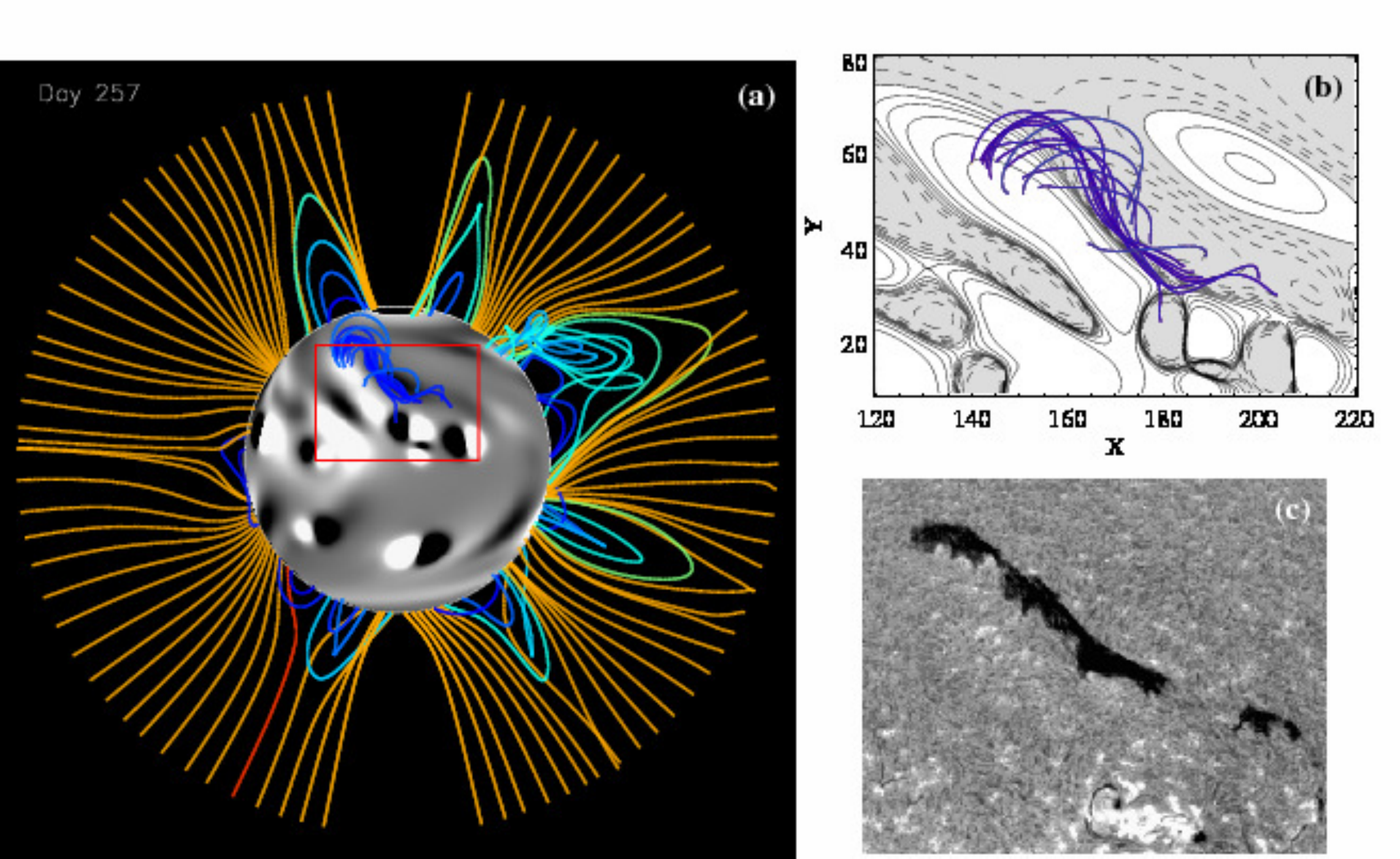}
\caption{Example of the comparison of theory and observations
performed by \cite{2008SoPh..247..103Y}.  (a)~Magnetic field
distribution in the global simulation after 108~days of evolution,
showing highly twisted flux ropes, weakly sheared arcades, and near
potential open fields. On the central image white/black represents
positive/negative flux. (b)~Close up view of a dextral flux rope lying
above a PIL within the simulation. (c)~BBSO H$\alpha$ image of the
dextral filament observed at this location. }
\label{fig:mackay_extra2}
\end{figure}}

\epubtkImage{mackay_fig1.png}{%
\begin{figure}[htb]
\centering\includegraphics[scale=0.3]{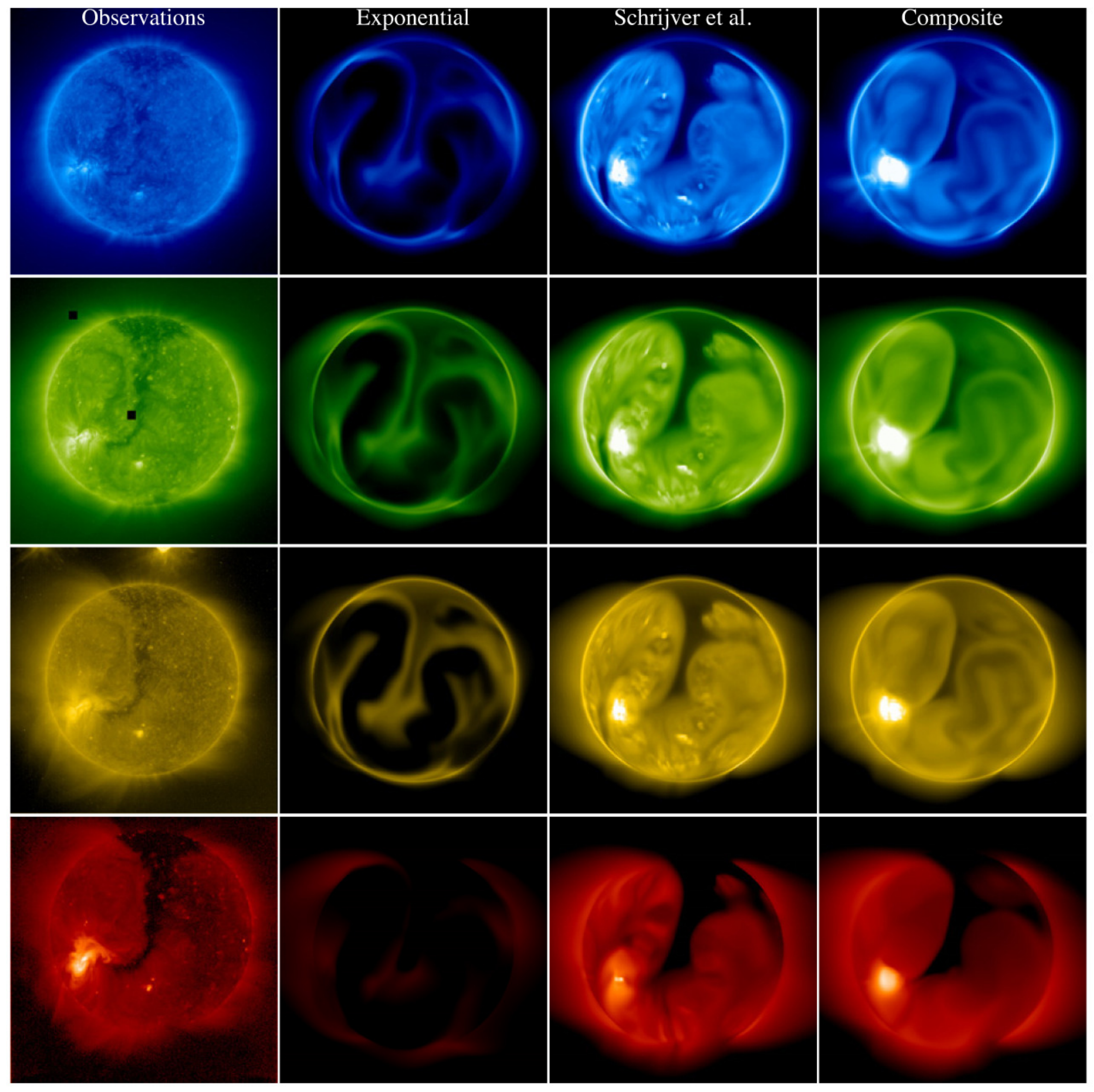}
\caption{Comparison of observations and simulated emissions from a
  global steady-state MHD model. The observations are shown in the
  first column, while the emissions due to different coronal heating
  profiles are shown in the other columns. Image reproduced by
  permission from Figure~8 of~\cite{2009ApJ...690..902L}, copyright by
  AAS.}
\label{fig:globalmhd}
\end{figure}}

\subsection{Plasma emission}
\label{sec:mhd}  
\label{sec:mhdss}

Although simple models of the coronal plasma emission may be derived by populating PFSS magnetic field lines with hydrostatic atmospheres \citep{2002MNRAS.336.1364J}, self-consistent models of plasma density and temperature require full-MHD models (Section~\ref{sec:mhdfull}). Indeed, a key emphasis of global MHD models has been direct comparison of the steady state 
solutions with either white-light \citep{2010AA...513A..45R} or multi-spectral EUV and 
X-Ray observations \citep{2009ApJ...690..902L,2010ApJ...712.1219D}.
Early global MHD models using a simplified polytropic energy equation 
were able to produce a good representation of the Suns large-scale magnetic field 
\citep{1999PhPl....6.2217M}, however they were unable to reproduce realistic emission 
profiles in EUV and X-rays. This was attributed to the fact that they did not produce 
a sufficiently high density and temperature contrast between open and closed field regions. 
To improve the models, a more realistic energy equation has been incorporated, including the effects 
of thermal conduction, radiative losses, and coronal heating, along 
with modeling the upper chromosphere and transition region 
\citep{2001ApJ...546..542L,2009ApJ...690..902L,2010ApJ...712.1219D}.

Figure~\ref{fig:globalmhd} shows a comparison of the global MHD model of 
\cite{2009ApJ...690..902L} with observations for CR 1913 (August\,--\,September
1996). 
The left hand column shows three EUV pass bands (171, 195, 284~\AA) along with a Yohkoh/SXT X-ray 
image. The other three columns show synthetic emission profiles constructed 
from density and temperature distributions in the global MHD simulation. 
Each column uses exactly the same MHD model: the only difference is the form of coronal heating applied ($H_{\mathrm{ch}}$ in Equation~\eqref{eqn:ent}). While each solution produces roughly the same magnetic structure, the most
important factor in reproducing the emissions is the coronal heating. The second column, which uses an exponential form falling off with height, gives the worst comparison. 
When spatially varying heating is applied (either from \citealp{2004ApJ...615..512S} or 
the composite form of \citealp{2009ApJ...690..902L}), the simulation captures many of the 
features of the observed emission. Similar results were also found in the paper
of \citet{2010ApJ...712.1219D}. In an alternative comparison, \cite{2010AA...513A..45R} 
compared the output from the global MHD model shown in Figure~\ref{fig:globalmhd} with white light
eclipse observations from 1~August 2008. To compare the model and observations, the authors
simulated the white light emission of the corona by integrating the electron density along the line-of-sight.
The results of the comparison can be seen in Figure~5 of \cite{2010AA...513A..45R} where the model
successfully reproduced the overall shape of the corona along with the shape and location of three helmet
streamers.

\clearpage

\section{Conclusions}
\label{sec:con}

In this review, our present day understanding of global magnetic fields on the Sun and other
stars has been discussed from both observational and theoretical viewpoints. For the Sun, we now have 
long-running data sets of global magnetic activity. For other stars, we are 
just beginning to learn of the wide variety of magnetic morphologies that exist. In terms of 
theoretical models, recent years have seen a significant advance in global modeling techniques. 
Global magnetic fields may be modeled under the nonlinear force-free approximation for long periods 
of time, or for short periods of time using highly detailed
MHD models. A key feature of these models is that they have been validated through various
comparisons with observations. While our understanding has significantly increased over the last
decade, there are several immediate outstanding issues. Five of these are discussed below, where 
the list is not exhaustive and the order in no way reflects their importance.

\begin{enumerate}
\item While we have a detailed understanding of the normal magnetic field component on the solar photosphere, 
the same is
not true for the vector field. Between SDO and SOLIS we now have daily full disk vector magnetograms. Analysis
of these will gain us an understanding of the emergence and transport of vector fields across the Sun,
as well as the origin and evolution of magnetic helicity, a key component in eruptive phenomena.
Additionally, vector magnetic field measurements are need not just in strong field regions, but also in weak field regions.

\item For theoretical models, regular observations of the vector fields mean that more
constraints can be applied to input data of bipole emergences that are used to drive these models. New techniques must be developed
for the incorporation of vector data into simulations. Only through this can theory and observations 
be truly unified. 

\item A number of inconsistencies with the application of surface flux transport models need to be resolved. These 
relate to the use of additional decay terms, variations in the rate or profile of meridional flow,
and the exact tilt angle applied for new emerging bipoles. Currently various combinations of all of these have
been shown (when coupled with different coronal models) to give similar results. It would be useful
for the various authors to resolve these issues through comparing results from their codes, driven by the
variety of input data sets available. In addition, more constraints on the input data are required from 
observations. 

\item While our understanding of stellar magnetic fields has greatly increased, observations are
still sporadic. Long term data sets of individual stars across different spectral types
are required. From this we may deduce different magnetic flux emergence and transport parameters
which are critical for the next generation of dynamo models.

\item Global time-dependent MHD models with evolving lower boundary conditions must be
developed, to provide a self-consistent model for the evolution and interaction of 
magnetic fields with plasma.

\end{enumerate}

These issues may be addressed with the current suite of ground and space-based observatories along with
developing new theoretical techniques. A key
element in achieving this goal is the increasing computing power available both for real-time data reduction 
and for theoretical modeling. 
Future revisions of the article will hopefully describe answers to some of these questions.  Along with answering 
these questions, the scope
of the review in future revisions will also be increased to include additional topics such as: the connection 
between the Sun and Heliosphere, 
observations of coronal and solar prominence magnetic fields, global eruptive models and a fuller discussion of
the consequences of observational limitations on the models which directly employ observations.

\section{Acknowledgments}
\label{section:acknowledgements}

DHM would like to thank STFC, the Leverhulme Trust and European Commission's 
Seventh Framework Programme (FP7/2007-2013) under the grant agreement SWIFF 
(project no.~263340, http://www.swiff.eu) for their financial
support. ARY thanks STFC for financial support while at the University
of Dundee. The authors would also like to thank the two anonymous referees for their positive comments
on the manuscript, along with Vic Gaizauskas and Ineke De Moortel for providing comments on the first draft.

\newpage

\bibliography{refs}

\end{document}